\documentclass[aip,jcp,reprint,amsmath,amssymb,floatfix,citeautoscript]{revtex4-1}
\usepackage{cancel}
\usepackage{amsmath}
\usepackage{mathtools}
\usepackage{physics}
\usepackage{graphicx}
\usepackage{amssymb}
\usepackage{amsthm}
\usepackage{bm}
\usepackage{dcolumn}
\usepackage{braket}
\usepackage{longtable}
\usepackage{ragged2e}
\usepackage{txfonts}
\usepackage[version=3]{mhchem} 
\usepackage[T1]{fontenc}       
\usepackage{pstricks}
\usepackage{siunitx}
\usepackage[colorlinks=true,allcolors=blue]{hyperref}
\usepackage[capitalise]{cleveref}
\usepackage{multirow}
\usepackage{algorithm}         
\usepackage[noend]{algpseudocode}     
\usepackage{booktabs}
\usepackage[flushleft]{threeparttable}

\usepackage{lipsum} 

\let\tr\undefined
\newcommand*\tr[1]{\textrm{#1}}

\definecolor{blackpink}{RGB}{200, 64, 200}
\newcommand*\rvv[1]{{#1}}

\newcommand*\Ha{$E_{\tr{h}}$}

\allowdisplaybreaks

\begin{document}

\title
{Correlation-consistent Gaussian basis sets for solids made simple}

\author{Hong-Zhou Ye}
\email{hzyechem@gmail.com}
\affiliation
{Department of Chemistry, Columbia University, New York, New York 10027, USA}
\author{Timothy C. Berkelbach}
\email{tim.berkelbach@gmail.com}
\affiliation
{Department of Chemistry, Columbia University, New York, New York 10027, USA}
\affiliation
{Center for Computational Quantum Physics, Flatiron Institute, New York, New York 10010, USA}

\begin{abstract}
    The rapidly growing interest in simulating condensed-phase materials using quantum chemistry methods calls for a library of high-quality Gaussian basis sets suitable for periodic calculations.
    Unfortunately, most standard Gaussian basis sets commonly used in molecular simulation show significant linear dependencies when used in close-packed solids, leading to severe numerical issues that hamper the convergence to the complete basis set (CBS) limit, especially in correlated calculations.
    In this work, we revisit Dunning's strategy for construction of correlation-consistent basis sets and examine the relationship between accuracy and numerical stability in periodic settings.
    We find that limiting the number of primitive functions avoids the appearance of problematic small exponents while still providing smooth convergence to the CBS limit.
    As an example, we generate double-, triple-, and quadruple-zeta correlation-consistent Gaussian basis sets for periodic calculations with Goedecker-Teter-Hutter (GTH) pseudopotentials.
    Our basis sets cover the main-group elements from the first three rows of the periodic table.
    \rvv{Especially for atoms on the left side of the periodic table, our basis sets are less diffuse than those used in molecular calculations.}
    We verify the fast and reliable convergence to the CBS limit in both Hartree-Fock and post-Hartree-Fock (MP2) calculations, using a diverse test set of $19$ semiconductors and insulators.
\end{abstract}

\maketitle


\section{Introduction}
\label{sec:introduction}

    Recent years have witnessed a rapid growth of interest in understanding condensed-phase materials using quantum chemistry methods \cite{Marsman09JCP,Muller12PCCP,DelBen12JCTC,DelBen13JCTC,Booth13Nature,Yang14Science,McClain17JCTC,Gruber18PRX,Zhang19FM,Wang20JCTC,Lau21JPCL,Lange21JCP,Wang21JCTC},
    especially those beyond density functional theory \cite{Hohenberg64PR,Kohn65PR} (DFT) with local and semi-local exchange-correlation functionals \cite{Perdew96PRL,Perdew08PRL}.
    The non-local exchange and the many-body electron correlation in the quantum chemistry methods promise many advantages, such as systematic improvability \cite{McClain17JCTC} and the ability to describe dispersion interactions \cite{Jeziorski94CR,Sinnokrot02JACS,Sherrill09JPCA} and strong electron correlations \cite{Zheng17Science,Li19NC}, but are also computationally demanding, especially when using a plane-wave (PW) basis set due to the slow convergence with the number of virtual bands~\cite{Marsman09JCP,Gruneis10JCP,Gruneis11JCTC,Shepherd12PRB,Jiang16PRB,Booth16JCP,Morales20JCP,Callahan21JCP,Wei21JCTC}.

    Atom-centered Gaussian basis sets are most popular in molecular quantum chemistry \cite{Hill13IJQC}, where the correlation-consistent basis sets \cite{Dunning89JCP} allow systematic convergence to the complete basis set (CBS) limit in correlated calculations. \cite{Feller11JCP,Neese11JCTC,Zhang13NJP,Varandas21PCCP}
    However, the properties that define a good basis set for molecules are not the same as those for periodic solids.
    For example, the standard Gaussian basis sets \cite{Feller96JCC,Weigend03JCP,Weigend05PCCP}, often optimized on free atoms, contain relatively diffuse functions that are needed to correctly describe the wavefunction in distant regions of space.
    These diffuse functions cause significant linear dependencies when used in periodic calculations \cite{Klahn77IJQC,VandeVondele07JCP,Peintinger13JCC,VilelaOliveira19JCC}, leading to numerical instabilities in the self-consistent field (SCF) calculations \cite{Roothaan51RMP}.
    While this SCF convergence issue can sometimes be solved by discarding diffuse primitives in the basis set \cite{Peintinger13JCC} or by canonical orthogonalization \cite{Lee21JCP}, these modifications hinder reproducibility, cause discontinuities in potential energy surfaces, and degrade the quality of virtual orbitals, which affects subsequent correlated calculations.
    More importantly, the linear dependency problem is worse for larger basis sets, preventing convergence of a periodic calculation to the CBS limit.

    One way to mitigate the linear dependency issue is re-optimizing the Gaussian exponents $\zeta_i$ and contraction coefficients $c_i$ of existing basis sets based on a cost function such as \cite{VandeVondele07JCP,Daga20JCTC,Li21JPCL,Zhou21JCTC,Neufeld21TBP}
    \begin{equation}    \label{eq:molbulkopt}
        \Omega(\{\zeta_i,c_i\}; \gamma)
            = E(\{\zeta_i,c_i\}) + \gamma \log \mathrm{cond}\mathbf{S}(\{\zeta_i,c_i\}),
    \end{equation}
    the minimization of which trades some of the energy $E$ for a lower condition number of the basis set overlap matrix $\mathbf{S}$ to the extent controlled by the developer-selected parameter $\gamma > 0$.
    Such a cost function can be minimized on a paradigmatic system that exhibits the linear dependency of concern \cite{VandeVondele07JCP,Li21JPCL} or on each system under study \cite{Daga20JCTC,Zhou21JCTC,Neufeld21TBP}, and recent works have demonstrated the success of such approaches for producing Gaussian basis sets with better behavior \cite{Li21JPCL,Daga20JCTC,Zhou21JCTC,Neufeld21TBP}.
    However, aside from the extra cost associated with frequent basis set reoptimization, such approaches obviously hinder---or forfeit---transferability and reproducibility.

    In this work, we take a different approach to constructing Gaussian basis sets for periodic systems, which are designed to be universal and transferable, by revisiting Dunning's strategy for generating correlation-consistent Gaussian basis sets \cite{Dunning89JCP,Woon93JCP}.
    The key modification needed for extended systems is found to be restricting the size of the valence basis to reach a balance between the accuracy and the numerical stability of a basis set.
    Our strategy is general and applies to both all-electron and pseudopotential-based calculations;
    as a specific example, we use the Goedecker-Teter-Hutter (GTH) family of pseudopotentials \cite{Goedecker96PRB,Hartwigsen98PRB} optimized for Hartree-Fock \cite{HutterPP} (HF) calculations and generate correlation-consistent Gaussian basis sets up to the quadruple-zeta (QZ) level for the main-group elements from the first three rows of the periodic table.
    The resulting GTH-cc-pV$X$Z ($X = $ D, T, and Q) basis set series show fast convergence to the CBS limit (verified using a PW basis with the same pseudopotential) on the bulk properties of $19$ semiconductors calculated at both mean-field (HF) and correlated (second-order M\o{}ller-Plesset perturbation theory \cite{Moller34PR}, MP2) levels.

    This article is organized as follows.
    In \cref{sec:methodology}, we review Dunning's original scheme for generating correlation-consistent basis sets and provide a high-level description of our adaptation for periodic systems, using the GTH-cc-pV$X$Z basis set of carbon as an illustrative example.
    In \cref{sec:computational_details}, we describe the computational details of the basis optimization and the numerical tests.
    In \cref{sec:results_and_discussion}, we evaluate the quality of the GTH-cc-pV$X$Z basis sets on a variety of bulk properties by comparing to results obtained with a PW basis.
    In \cref{sec:conclusion}, we conclude this work by pointing out future directions.

\section{Methodology}
\label{sec:methodology}


    \subsection{Correlation consistent basis sets}
    \label{subsec:correlation_consistent_basis_sets}

    We begin with a brief review of Dunning's original approach to constructing the cc-pV$X$Z basis set series for the main-group elements, \cite{Dunning89JCP} which our strategy closely follows.
    A cc-pV$X$Z basis set consists of a valence basis and a set of polarization functions.
    The valence basis has $s$ and $p$ primitive orbitals (only $s$ for hydrogen and helium) whose exponents are determined by minimizing the HF ground state energy of a free atom.
    These optimized primitive orbitals are then contracted with coefficients obtained from spherically averaging the atomic HF orbitals.
    The most diffuse one, two, etc.~primitive orbitals are freed from the contraction for DZ, TZ, etc.~to better describe the electron density in the bonding region of a molecule.
    The size of the valence basis (i.e.,~the number of primitive $s$ and $p$ orbitals) is typically chosen by the desired accuracy of the atomic HF energy.

    The polarization functions are primitive orbitals of $d$ angular momentum or higher ($p$ or higher for hydrogen and helium) whose exponents are determined by minimizing the correlation energy of a free atom.
    The rule of correlation consistency---an empirical observation first made by Dunning \cite{Dunning89JCP} and later confirmed by other \cite{Woon93JCP,Balabanov05JCP,Prascher11TCA}---states that the increase in the magnitude of the correlation energy $|\Delta E_\mathrm{c}|$ obtained by adding the $n$th polarization function of angular momentum $l$ is roughly equal to that of adding the $(n-1)$th polarization function of angular momentum $l+1$.
    For this reason, the polarization functions are added in groups, $1d$ for DZ, $2d1f$ for TZ, $3d2f1g$ for QZ, etc., and the correlation energies obtained with the cc-pV$X$Z series can often be extrapolated to the CBS limit using simple functional forms \cite{Feller11JCP,Neese11JCTC}.

    \subsection{The linear dependency problem}
    \label{subsec:lindep_problem}

    \begin{figure}[!t]
        \centering
        \includegraphics[scale=0.8]{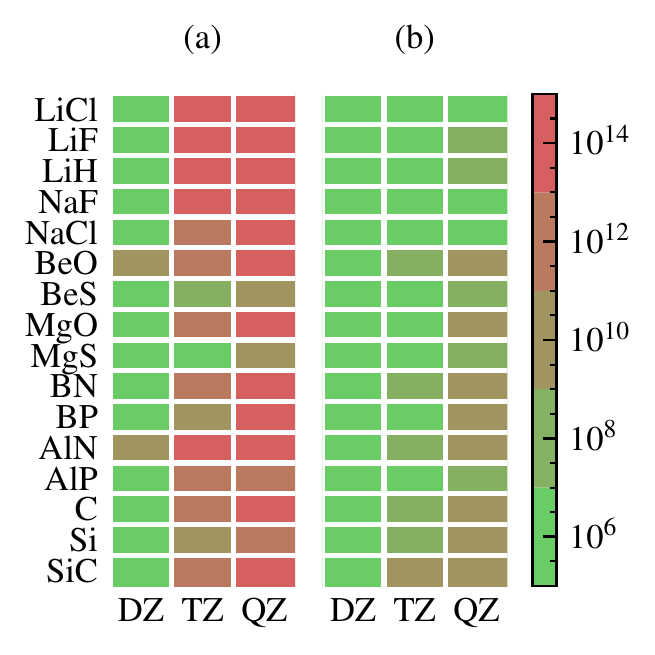}
        \caption{Condition number of (a) the GTH-$X$ZVP (taken from the CP2K software package\cite{Kuhne20JCP}) and (b) the GTH-cc-pV$X$Z (this work) basis sets evaluated on $16$ three-dimensional bulk materials at their respective experimental lattice parameters using a $5\times5\times5$ $k$-point mesh (the maximum condition number from all $k$-points is plotted).
        For materials containing $s$-block elements, the condition numbers from the small-core and the large-core GTH-cc-pV$X$Z basis sets are comparable and the former is shown here.}
        \label{fig:condnum_cmp}
    \end{figure}

    The primary problem that must be solved for generating Gaussian basis sets for solids is the potential high linear dependency of the basis sets \cite{Klahn77IJQC,VandeVondele07JCP,Peintinger13JCC,VilelaOliveira19JCC}, which is particularly severe for three-dimensional solids.
    This is illustrated in \cref{fig:condnum_cmp}(a) for the original GTH-$X$ZVP basis sets \cite{VandeVondele05CPC} on $16$ bulk solids composed of the main-group elements from the first three rows.
    The GTH-$X$ZVP basis sets, which were first reported in Ref.~\citenum{VandeVondele05CPC} and are now distributed with the CP2K package \cite{Kuhne20JCP}, were constructed by combining a valence basis optimized on free atoms at the DFT level and polarization functions of $d$ angular momentum taken from the corresponding cc-pV$X$Z basis sets.
    In practice, a condition number higher than $10^{10}$ is found to be problematic in the manners discussed in~\cref{sec:introduction}.
    As a result, most solids listed in \cref{fig:condnum_cmp}(a) can only be studied at the DZ level when using the original GTH basis sets.
    The situation is similar for other all-electron \cite{Dunning89JCP,Woon93JCP,Weigend05PCCP} or pseudopotential-based basis sets \cite{Bennett17JCP,Bennett18JCP}, including the cc-pV$X$Z series \cite{Dunning89JCP,Woon93JCP}, as shown in Fig.~S1.
    We emphasize that this issue is mostly isolated to hard crystalline solids and is less severe for molecular solids or liquids.

    \subsection{Balancing accuracy and numerical stability}
    \label{subsec:balancing_acc_and_numstab}

    \begin{figure*}[!t]
        \centering
        \includegraphics[scale=0.8]{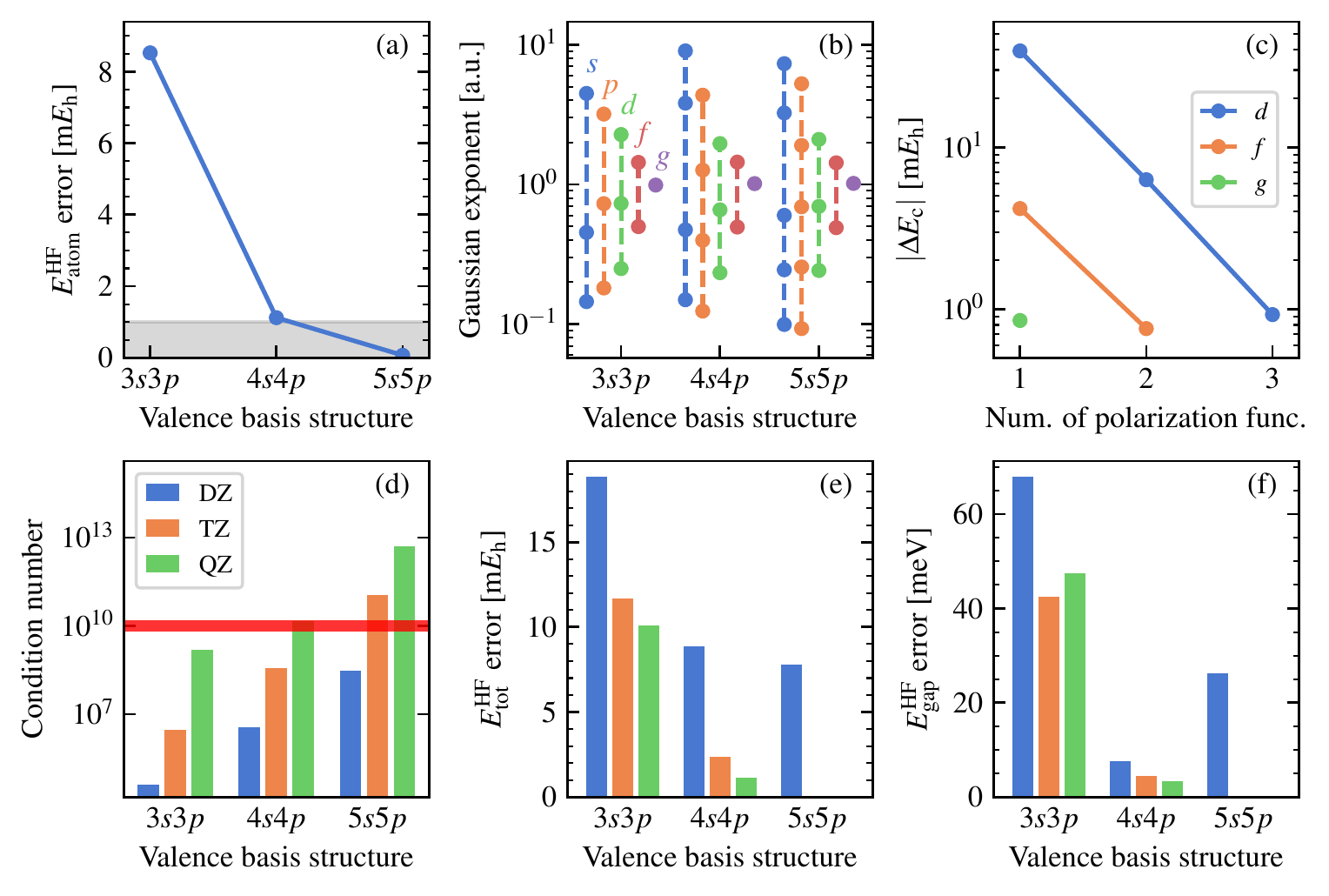}
        \caption{Correlation consistent basis sets generated for carbon in the GTH pseudopotential using three different valence bases.
        (a) Convergence of the atomic HF energy. The grey shaded area indicates an error below $1$ m\Ha{}.
        (b) Gaussian exponents of the QZ basis sets. Different colors label shells of different angular momentum as indicated by the text in the corresponding color.
        (c) Increment of the atomic CCSD correlation energy with the number of polarization functions in each angular momentum channel for the $4s4p$ valence basis. The $3s3p$ and $5s5p$ valence bases give virtually the same plot (not shown).
        (d) Condition number of the basis overlap matrix evaluated for bulk diamond at the experimental geometry (the maximum condition number from a $5\times5\times5$ $k$-point mesh is plotted).
        The red horizontal line highlights a condition number of $10^{10}$.
        (e) Error of the per-cell HF energy of bulk diamond evaluated using the Gaussian basis sets against a PW benchmark calculation.
        (f) Same as (e) for the HF band gap of diamond.}
        \label{fig:C_acc_vs_cond}
    \end{figure*}

    In our approach to basis set design for solids, we control the linear dependency by limiting the size of the valence basis, being careful not to introduce large basis set incompleteness errors.
    In this section, we use the carbon element with a GTH pseudopotential as an example to discuss how a balance between accuracy and numerical stability can be reached.
    We postpone a discussion of computational details to \cref{sec:computational_details}.

    \Cref{fig:C_acc_vs_cond}(a) shows the error in the atomic HF energy of carbon with three optimized valence bases of increasing size (i.e., number of primitives): $3s3p$, $4s4p$, and $5s5p$.
    A relatively small basis of $4s4p$ achieves an error of about 1~m\Ha{}, and that of $5s5p$ is already below $0.1$~m\Ha{}.
    For each of the three valence bases, we generate correlation-consistent DZ, TZ, and QZ basis sets, by optimizing the polarization functions based on a correlated calculation.
    The optimized exponents of the QZ primitives are shown in \cref{fig:C_acc_vs_cond}(b). Focusing on the $s$ primitives, we note that the \textit{largest} exponent splits into two from $3s$ to $4s$, but the \textit{smallest} exponent splits into two from $4s$ to $5s$. Only the latter is consistent with the conventional ``split-valence'' picture. We will see that this split-valence structure yields unacceptably high condition numbers in solids, but is not necessary for accurate predictions.
    Despite the difference in the underlying valence basis, the optimized exponents for the polarization functions share a similar structure [\cref{fig:C_acc_vs_cond}(b)] and all exhibit perfect correlation consistency as shown explicitly for the $4s4p$ case in \cref{fig:C_acc_vs_cond}(c).

    To test the applicability and the performance of the nine correlation-consistent bases obtained above in periodic calculations, we consider bulk diamond with its experimental lattice constant.
    The condition numbers plotted in \cref{fig:C_acc_vs_cond}(d) exhibit a quick and monotonic increase with both the zeta-level and the size of the valence basis.
    As a result, we observed convergence issues in the SCF calculations using the $5s5p$-derived TZ and QZ basis sets, consistent with the high condition numbers of the two basis sets (greater than $10^{10}$ as indicated by the red line), which is due to the presence of exponents near $0.1$ or below for this system.

    For the remaining seven basis sets (DZ to QZ for $3s3p$ and $4s4p$ and DZ for $5s5p$), we used HF and MP2 to calculate various structural and energetic properties of diamond, which were then compared with benchmark results obtained using a PW basis.
    For most of these properties, the performance of the correlation-consistent basis sets at the same zeta-level shows a weak dependence on the underlying valence basis (Figs.~S2 and S3).
    However, the $4s4p$ family is a clear winner on the more sensitive properties, including the HF total energy and the HF band gap as shown in \cref{fig:C_acc_vs_cond}(e-f); large residual errors persist in the $3s3p$-derived basis sets, even at the QZ level, and little improvement is gained in the $5s5p$-derived basis sets (at least at the DZ level which is the only one we can test due to their high linear dependencies).

    The results in this section reveal the strong effect of the choice of a valence basis on the accuracy and the numerical stability of the resulting correlation-consistent basis sets.
    In particular, we see that limiting the size of the valence basis can preclude the appearance of problematic diffuse functions without significantly compromising the accuracy of calculated properties.
    In the next section, we extend the strategy used here to obtain correlation-consistent basis sets for all main-group elements from the first three rows of the periodic table.

    \subsection{The GTH-cc-pV$X$Z basis sets}
    \label{subsec:gthccpvxz}

    In the following subsections, we describe the detailed construction of our GTH-cc-pV$X$Z basis sets ($X = $ D, T, and Q) for the first three rows of the periodic table.
    For selected elements (\textit{vide infra}), we also construct basis sets augmented by diffusion functions (GTH-aug-cc-pV$X$Z).
    All of our basis sets are available for download in an online repository~\cite{ccgto_github} and full details of their primitive and contracted structure is given in Table S1 and Figs.~S4 and S5.

    \subsubsection{The valence basis}
    \label{subsubsec:gth_valence_basis}

    The scheme for choosing the optimal valence basis for the carbon atom with a GTH pseudopotential, presented in \cref{subsec:balancing_acc_and_numstab}, can be made general.
    Based on atomic calculations, we generate candidate correlation-consistent basis sets using valence bases of multiple sizes, which are then tested on a few reference materials.
    The final basis set is then chosen as the one that remains numerically stable while predicting bulk properties that are converged with the size of the valence basis (or as converged as possible before large linear dependencies arise).
    \rvv{In contrast to Dunning's original approach, we use these primitives for all zeta levels, with the final valence basis differing only in the number of uncontracted functions (however, see \ref{subsubsec:contraction_valence_orbitals} for a modification of this procedure for group VI to VIII elements).}
    We emphasize that the reference periodic system serves only as a \emph{guide} for choosing an appropriate valence basis and is not used in the optimization of any parameters, unlike in previous works based on \cref{eq:molbulkopt} \cite{VandeVondele07JCP,Daga20JCTC,Li21JPCL,Zhou21JCTC,Neufeld21TBP}.
    As we will see in the numerical results (\cref{sec:results_and_discussion}), our scheme maintains the important atomic electronic structure of a basis set, which is crucial for its transferability and high accuracy.

    The structure of the valence basis determined this way for all elements from the first three rows of the periodic table is summarized in \cref{tab:valence_basis}.
    The reference systems are chosen to be simple semiconductors and insulators formed by these elements.
    The bulk properties being monitored include the equilibrium lattice constant and bulk modulus evaluated at both the HF and the MP2 levels and the HF band gap at equilibrium geometry, all evaluated with the Brillouin zone sampled by a $3\times3\times3$ $k$-point mesh.
    The $s$-block elements, Li, Be, Na, and Mg, can be simulated using either a large-core pseudopotential or a small-core pseudopotential (which differ according to the treatment of core/semi-core electrons), and we present optimized basis sets for them separately.
    Furthermore, for these $s$-block elements, the exponents of the valence $p$ orbitals cannot be determined in the usual way because these orbitals are unoccupied in the atomic HF ground state.
    In the literature, these exponents are often determined by minimizing the HF energy of the corresponding $s \to p$ valence excited state \cite{Prascher11TCA}.
    We found that this approach leads to valence $p$ orbitals that are too diffuse and cause significant linear dependencies in the oxides and fluorides of these elements using the QZ basis sets.
    Therefore, we optimize the valence $p$ orbitals at the correlated level in the same way as in the determination of the polarization functions, described more in the next section.
    We verified that the DZ and TZ basis sets obtained from both schemes give very similar numerical results for all properties tested in \cref{sec:results_and_discussion}.

\begin{figure}[t]
        \centering
        \includegraphics[scale=0.35]{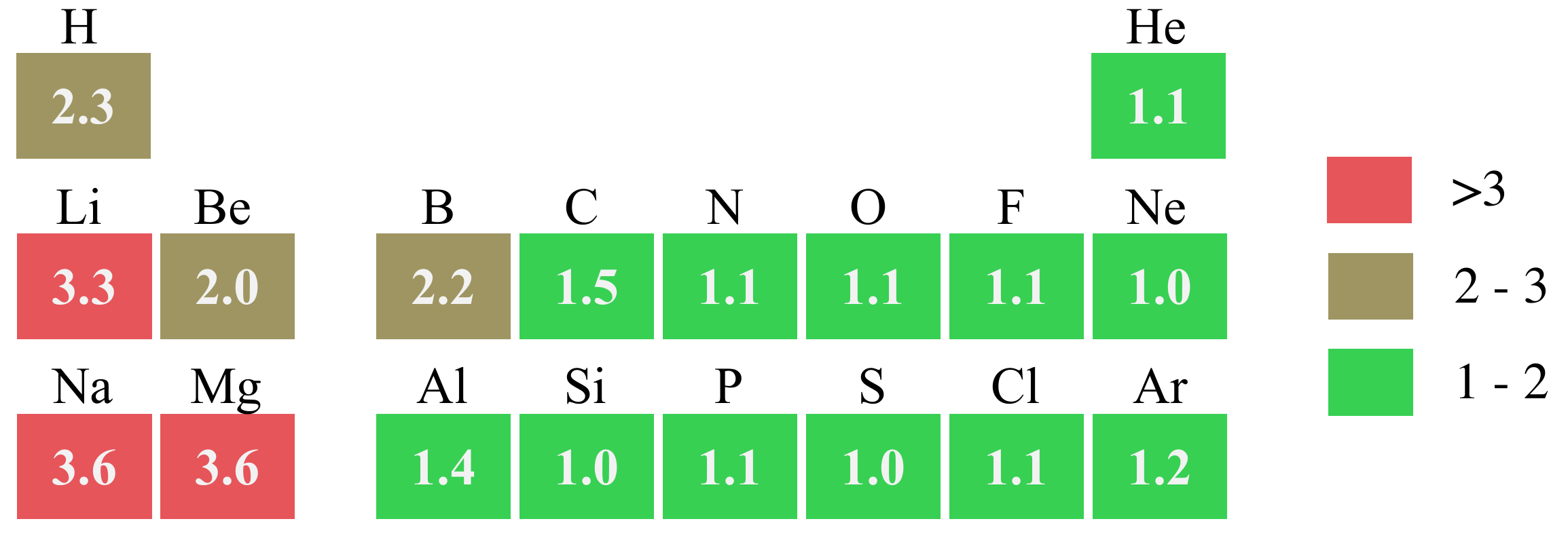}
        \caption{Compactness of the valence basis developed in this work, defined as the ratio of the smallest exponent in our basis to that in the all-electron cc-pV$5$Z basis set for the same element.
        \rvv{For $s$-block elements, the compactness of the valence basis optimized for the small-core and the large-core pseudopotentials is comparable, and we only show the results for the former.}}
        \label{fig:sp_expn_rat}
    \end{figure}

    The size of the valence basis is found to correlate with the hardness of the underlying pseudopotentials as reported in previous work \cite{Li21JPCL} but is otherwise smaller than the valence bases in the original GTH-TZVP and QZVP basis sets.
    For this reason, the condition numbers of the new basis sets are significantly reduced compared to the original GTH-$X$ZVP series as shown in \cref{fig:condnum_cmp}(b).
    \rvv{This can also be seen from the \emph{compactness} of our valence basis (\cref{fig:sp_expn_rat}), defined as the ratio of the smallest exponent in our valence basis to that in the corresponding all-electron cc-pV$5$Z basis set (geometric mean is taken in case of multiple angular momentum channels).}
    \rvv{From this perspective, the constraint on the size of the valence basis is strongest for the $s$-block metals (compactness $> 3$), which explains the relatively large error of the atomic HF energy for these elements, but gradually relaxed for elements of higher group numbers (compactness $\approx 1$).}
    As we will see by thorough numerical tests in \cref{sec:results_and_discussion}, this way of constraining the size of the valence basis does not degrade the performance of the full correlation-consistent basis sets in bulk calculations.
    \rvv{The compactness of the valence basis in \cref{fig:sp_expn_rat} also provides a practical guide to constructing valence bases of similar quality for other nuclear potentials (including the full Coulomb potential for all-electron calculations).}

    \begin{table}[t]
        \centering
        \caption{The active electrons (not covered by the pseudopotential), the valence basis structure (i.e.,~number of primitives), the errors of the atomic HF energy (in m\Ha{}, \rvv{evaluated as the energy difference from a sufficiently large valence basis}), and the reference systems on which the basis linear dependency and accuracy are monitored for the GTH-cc-pV$X$Z basis sets of all elements studied in this work.
        For the $s$-block elements, the valence bases are listed for both the large-core and the small-core pseudopotentials.
        The valence $p$ orbitals of the $s$-block elements are obtained in a different manner (see the discussion at the end of \cref{subsubsec:gth_valence_basis}) and denoted by a "+" sign in the table.}
        \label{tab:valence_basis}
        \begin{ruledtabular}
        \begin{tabular}{lllll}
            Element & Active electrons & Valence basis & HF error & Ref.~sys.~ \\
            \hline
            H & $[1s^1]$ & $4s$ & $0.19$ & LiH \\
            He & $[1s^2]$ & $5s$ & $0.05$ & solid He\\
            Li
                & $[2s^1]$ & $2s+2p$ & $1.24$ & LiH, LiF, LiCl \\
                & $[1s^22s^1]$ & $4s+\rvv{4}p$ & $2.39$ & LiH, LiF, LiCl \\
            Be
                & $[2s^2]$ & $\rvv{3}s+3p$ & $\rvv{0.12}$ & BeO, BeS \\
                & $[1s^22s^2]$ & $\rvv{5}s+4p$ & $\rvv{1.53}$ & BeO, BeS \\
            B & $[2s^22p^1]$ & $3s3p$ & $\rvv{8.15}$ & BN, BP \\
            C & $[2s^22p^2]$ & $4s4p$ & $1.13$ & diamond, SiC \\
            N & $[2s^22p^3]$ & $5s\rvv{5}p$ & $\rvv{0.19}$ & BN, AlN \\
            O & $[2s^22p^4]$ & $5s5p$ & $0.41$ & BeO, MgO \\
            F & $[2s^22p^5]$ & $5s5p$ & $0.72$ & LiF, NaF \\
            Ne & $[2s^22p^6]$ & $6s6p$ & $0.16$ & solid Ne \\
            Na
                & $[3s^1]$ & $2s+2p$ & $0.93$ & NaF, NaCl \\
                & $[2s^22p^63s^1]$ & $5s5p+1p$ & $2.44$ & NaF, NaCl \\
            Mg
                & $[3s^2]$ & $2s+1p$ & $4.32$ & MgO, MgS \\
                & $[2s^22p^63s^2]$ & $4s4p+1p$ & $17.27$ & MgO, MgS \\
            Al & $[3s^23p^1]$ & $3s2p$ & $0.62$ & AlN, AlP \\
            Si & $[3s^23p^2]$ & $4s4p$ & $0.05$ & Si, SiC \\
            P & $[3s^23p^3]$ & $4s4p$ & $0.13$ & AlN, AlP \\
            S & $[3s^23p^4]$ & $5s4p$ & $0.18$ & BeS, MgS \\
            Cl & $[3s^23p^5]$ & $5s4p$ & $0.24$ & LiCl, NaCl \\
            Ar & $[3s^23p^6]$ & $4s4p$ & $0.76$ & solid Ar
        \end{tabular}
        \end{ruledtabular}
    \end{table}

    \subsubsection{The polarization functions}
    \label{subsubsec:gth_polarization_functions}

    We applied Dunning's scheme for determining the polarization functions up to QZ (i.e.,~$3d2f1g$) without modifications to the elements from group III to VIII.
    However, we used coupled-cluster theory with single and double excitations (CCSD)~\cite{PurvisIII82JCP} instead of the more common configuration interaction with single and double excitations (CISD)~\cite{Dunning89JCP,Woon93JCP,Prascher11TCA}, for the calculation of the correlation energy.
    The increment of the correlation energy with the number of polarization functions follows the rule of correlation consistency for all these elements.
    Plots similar to \cref{fig:C_acc_vs_cond}(c) for other atoms are shown in Fig.~S6.

    The $s$-block elements (including hydrogen) need some special treatments.
    The hydrogen atom only has one electron and hence no correlation energy.
    We follow ref \citenum{Prascher11TCA} and minimize the correlation energy of a \ce{H2} molecule with the experimental bond length of $0.7414$ \AA{} \cite{Huber79Book}, which gives polarization functions showing good correlation consistency [Fig.~S6(a)].

    For $s$-block elements with small-core pseudopotentials, the $[1s^2]$ core electrons (for Li and Be) or the $[2s^2 2p^6]$ semi-core electrons (for Na and Mg) are frozen in the CCSD calculations in order to determine the polarization functions that only account for the valence electron correlation, following ref \citenum{Prascher11TCA}.
    With this choice, the same number of electrons are correlated when using either the large-core or the small-core pseudopotentials, leading to similar exponents for the optimized polarization functions for both pseudopotentials.
    For Li and Na with only one correlated valence electron, we apply a similar treatment as we did for the hydrogen atom and minimize the correlation energy of a \ce{Li2} molecule and a \ce{Na2} molecule with the respective experimental bond lengths of $2.673$ \AA{} and $2.303$ \AA{}. \cite{Huber79Book}
    However, we found significant linear dependencies for both elements in periodic calculations upon adding the second polarization function in each angular momentum channel.
    We thus choose a $1d$, $1d1f$, and $1d1f1g$ structure for the \emph{valence}-correlated DZ, TZ, and QZ polarization functions of these elements.
    For Be and Mg, the regular structure of polarization functions is kept, but deviation from ideal correlation consistency is observed, wherein the third $d$ function recovers much less correlation energy than the second $f$ function [Figs.~S6(e,f,o,p)].
    Similar observations have also been reported for these elements in the all-electron cases. \cite{Prascher11TCA}

    \rvv{Despite the use of the CCSD correlation energy and a pseudopotential, the exponents of the polarization functions in our GTH-cc-pV$X$Z basis sets in general agree very well with those in the all-electron cc-pV$X$Z basis sets (Figs.~S4 and S5).
    The few exceptions come from the third-row elements such as Al and Si, where our CCSD-optimized $3d$ set (for GTH-cc-pVQZ) shows a larger exponent splitting from the $2d$ set (for GTH-cc-pVTZ) than the corresponding $3d$ set in the all-electron cc-pVQZ basis sets.
    In these cases, we verified explicitly that very similar exponents are obtained by minimizing the CISD correlation energy instead.
    Thus, the observed difference between our GTH-cc-pV$X$Z series and the cc-pV$X$Z series for these elements is due to the pseudopotential.}

    \subsubsection{Contraction of valence orbitals}
    \label{subsubsec:contraction_valence_orbitals}

    In an all-electron cc-pV$X$Z basis set, the primitive orbitals in the valence basis describe both the core and the valence electrons of an atom, with those for the core being contracted in the way discussed in \cref{subsec:correlation_consistent_basis_sets} to reduce the computational cost.
    We follow this rule formally in constructing the GTH-cc-pV$X$Z basis sets \rvv{for most elements}, where the most diffuse one, two, and three primitive orbitals in each angular momentum channel of the valence basis are released from the contraction for $X = $ D, T, and Q.
    (In case of, e.g.,~three primitive orbitals in an angular momentum channel, TZ and QZ will have the same valence basis, which is fully uncontracted in that angular momentum channel.)
    \rvv{However, for group VI to VIII elements where} \rvv{the size of the valence basis is only weakly constrained (\cref{fig:sp_expn_rat}),} \rvv{the procedure above results in suboptimal performance, especially at the DZ level.
    For these elements, we instead augment the fully contracted valence basis with $1s1p$, $2s2p$, and $3s3p$ primitives ($s$-only for helium), determined separately from minimizing the atomic correlation energy (in the presence of the polarization functions), to make the DZ, TZ, and QZ basis sets.}

    The contraction coefficients determined from atomic HF orbitals might have limited transferability especially for group III to VIII elements and $s$-block elements with large-core pseudopotentials, because they have no core electrons.
    For this reason, we also generate valence-uncontracted DZ and TZ basis sets where the valence basis is made to match higher zeta-levels.
    For example, a GTH-cc-pV(T)DZ basis set has the polarization functions taken from GTH-cc-pVDZ and the valence basis from GTH-cc-pVTZ.
    We will see the importance of such valence uncontraction in \cref{sec:results_and_discussion} in the calculation of virtual bands.

    \subsubsection{\rvv{Extensions}}
    \label{subsubsec:gth_diffuse}

    \rvv{Like the all-electron cc-pV$X$Z series, our basis sets can be straightforwardly extended by core-valence correlating functions \cite{Woon95JCP,Peterson02JCP}, tight $d$ functions \cite{Dunning01JCP}, etc.}
    \rvv{In this work, we explore one such extension, namely the augmentation with diffuse functions}~\cite{Woon94JCP}, which may be appropriate for the simulation of molecular crystals \cite{DelBen12JCTC,Yang14Science} or surface phenomena \cite{Schimka10NM,Tsatsoulis17JCP,Lau21JPCL}.
    Particularly, we find that the noble gas solids, which are used as the reference materials for the evaluation of our noble gas basis sets, benefit substantially from augmentation with diffuse functions.
    We thus augment all Gaussian basis sets for the three noble gas elements used in this work by adding one diffuse function to each angular momentum channel.
    \rvv{Because of the low density of noble gas solids, the condition numbers of their overlap matrices are $10^{7}$ or less, even after augmentation.}
    The exponent of this augmentation function is chosen to be proportional to the exponent of the most diffuse function in the non-augmented basis: $\alpha_\tr{aug} = x \alpha_\tr{min}$, where $x$ is the analogous ratio of exponents in the all-electron aug-cc-pV$X$Z basis set for the same element. \cite{Woon94JCP}
    We name these basis sets ``GTH-aug-cc-pV$X$Z''.

\section{Computational details}
\label{sec:computational_details}


    The protocol for generating correlation-consistent basis sets described in \cref{sec:methodology} was followed, and all calculations are performed using the PySCF software package \cite{Sun18WIRCMS,Sun20JCP}.
    The spin-restricted (or spin-restricted open-shell) HF and the spin-unrestricted CCSD are used to optimize the valence basis and the polarization functions, respectively.
    For periodic calculations, the recently developed range-separated Gaussian density fitting \cite{Ye21JCPa,Ye21JCPb} (RSGDF) is used to handle the electron repulsion integrals.
    The density fitting auxiliary basis is an even-tempered Gaussian basis with a progression factor $\beta = 2.0$ (generated automatically by PySCF).
    Finite-size errors associated with the divergence of the HF exchange integral at $G=0$ are handled using a Madelung constant, as described in ref \citenum{Paier06JCP,Broqvist09PRB,Sundararaman13PRB}.
    The basis set parameters are optimized using the Nelder-Mead algorithm \cite{Nelder65CJ} (as implemented in the SciPy library \cite{Virtanen20NM}), which we found to give the same exponents as the Broyden-Fletcher-Goldfarb-Shanno \cite{Press92Book} or the conjugate gradient \cite{Hestenes52JRNBS} algorithms (as used in previous work \cite{Daga20JCTC,Zhou21JCTC}) in most cases, but to be more robust against local minima in more challenging situations.

    We assess the quality of the GTH-cc-pV$X$Z basis sets along with their valence-uncontracted counterparts at both HF and MP2 levels on a test set of $19$ three-dimensional bulk systems listed in Table S2 (the $16$ materials shown in \cref{fig:condnum_cmp} plus the solids of helium, neon, and argon).
    Results from the original GTH-DZVP basis sets, which show no linear dependency issues on these materials, will also be reported.
    For the three noble gas elements, we augment the GTH-DZVP basis with extra diffuse functions in the same manner as described in \cref{subsubsec:gth_diffuse}.
    For systems containing $s$-block elements that have small-core and large-core pseudopotentials, calculations in the GTH-cc-pV$X$Z family use the corresponding small-core and large-core basis sets that we developed, while those in the GTH-DZVP family use the small-core basis sets for both pseudopotentials, because large-core GTH-DZVP basis sets do not exist.

    The selected bulk properties include the cell energy ($E_{\tr{cell}}$), the cohesive energy ($E_{\tr{coh}}$), the band gap ($E_{\tr{gap}}$) and band structure (only at the HF level), and the equilibrium lattice constant ($a_0$) and bulk modulus ($B_0$).
    For $E_{\tr{tot}}$, $E_{\tr{coh}}$, and $E_{\tr{gap}}$, single-point calculations at experimental geometries are performed with the Brillouin zone sampled using a $5\times5\times5$ $k$-point mesh (evenly spaced and $\Gamma$-point included) without further extrapolation.
    The cohesive energy is counterpoise corrected for basis set superposition error.
    The band structure is obtained by performing individual single-point calculations using a $3\times3\times3$ $k$-point mesh shifted along a chosen $k$-point path.
    The $a_0$ and $B_0$ are obtained by scanning the lattice constant around the HF minimum and fitting the total energy curve to the Birch-Murnaghan equation of state \cite{Murnaghan44PNAS,Birch47PR}.
    A $3\times3\times3$ $k$-point mesh is used in all these calculations, except the results in \cref{tab:LiH_cf_PAW}, which were calculated using a $5\times 5\times 5$ $k$-point mesh to facilitate comparison with literature values.

    The errors in the above properties due to the Gaussian basis set incompleteness are determined by comparison to calculations with a PW basis made large enough to essentially achieve the CBS limit; the HF total energy, the HF band energy, and the MP2 energy are all converged to an accuracy better than $0.1$ meV/cell.
    We note that the $k$-point meshes used in all of our calculations are sufficiently large to eliminate finite-size effects in \textit{all} of the basis set errors and in \textit{most} (but not all) of the predicted properties.
    In our PW calculations, the HF exchange is calculated with the adaptively compressed exchange (ACE) operator. \cite{Lin16JCTC}
    To converge the MP2 correlation energy to the CBS limit, we extrapolate according to the asymptotic behavior~\cite{Shepherd12PRB}
    \begin{equation}    \label{eq:MP2_corr_nvir}
        E_{\tr{corr}}^{\tr{MP2}}(n_{\tr{vir}})
            = A n_{\tr{vir}}^{-1} + E_{\tr{corr}}^{\tr{MP2}}(\infty),
    \end{equation}
    where $n_{\tr{vir}}$ is the number of virtual bands per $k$-point.
    The proper range of $n_{\tr{vir}}$ for a safe extrapolation using \cref{eq:MP2_corr_nvir} is determined by monitoring the convergence of the properties calculated using the estimated $E_{\tr{c}}^{\tr{MP2}}(\infty)$.
    For the bulk systems composed of the elements from group III to V, well-converged estimates of $a_0^{\tr{MP2}}$ and $B_0^{\tr{MP2}}$ can be obtained by using $n_{\tr{vir}} = 350 \sim 400$ with uncertainties of about $0.1$ pm and $1$ GPa, respectively.
    Unfortunately, a much larger $n_{\tr{vir}}$, which is beyond the reach of our computational resources, is needed for the other bulk systems that contain the $s$-block or the noble gas elements and for all atomic calculations.
    Therefore, at the MP2 level, we are unable to evaluate basis set errors of all properties of these materials and the cohesive energy of all materials.
    However, we will study LiH with the small-core pseudopotential for Li as one such example of a difficult case.
    For this material, we will compare our results to MP2 calculations reported in literature and discuss how the Gaussian basis sets developed in this work can be leveraged to significantly improve the convergence of correlated calculations using a PW basis.

\section{Results and discussion}
\label{sec:results_and_discussion}


    \begin{figure}[!t]
        \centering
        \includegraphics[scale=0.8]{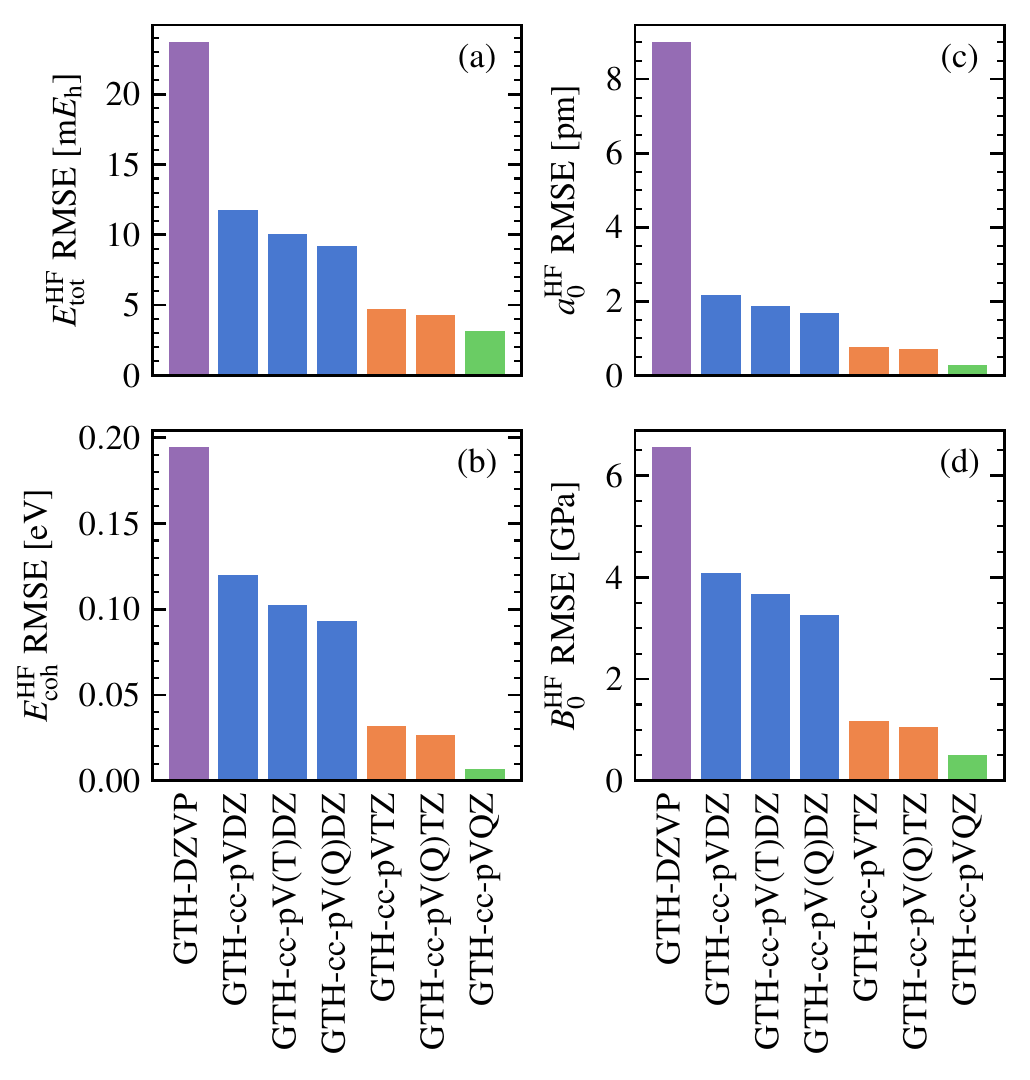}
        \caption{The root-mean-square error of (a) the cell energy, (b) the cohesive energy, (c) the equilibrium lattice constant, and (d) the equilibrium bulk modulus calculated at the HF level using different Gaussian basis sets for the $19$ bulk materials.}
        \label{fig:hf_totcoha0b0}
    \end{figure}

    \subsection{Occupied bands: HF ground-state properties}
    \label{subsec:occ_bands}

    \begin{figure*}[t]
        \centering
        \includegraphics[scale=0.8]{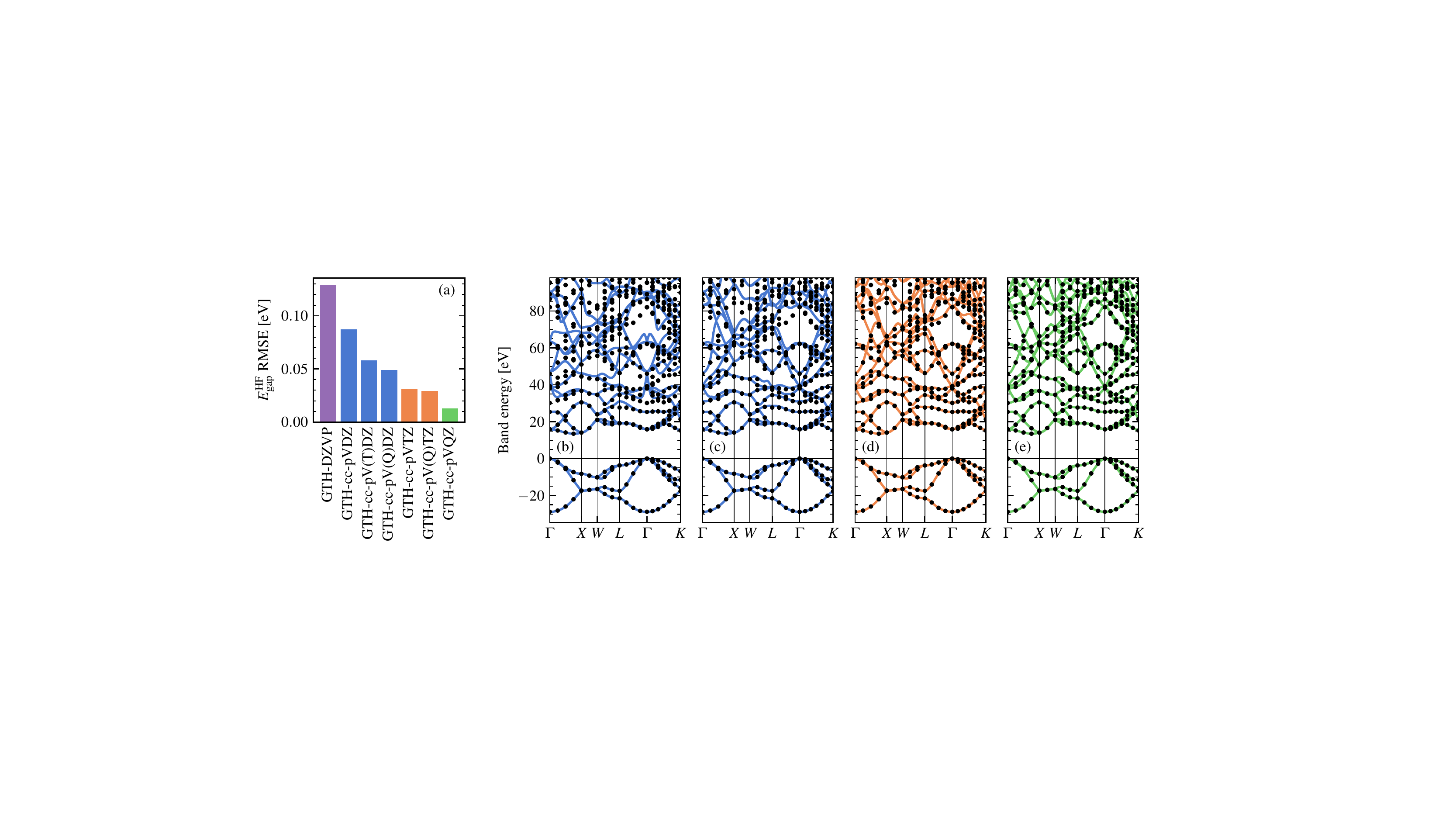}
        \caption{(a) The root-mean-square error of the HF band gap calculated using different Gaussian basis sets for the $19$ bulk materials.
        (b-e) HF band structure for diamond calculated using different Gaussian basis sets: (b) GTH-cc-pVDZ, (c) GTH-cc-pV(T)DZ, (d) GTH-cc-pVTZ, and (e) GTH-cc-pVQZ.
        The PW bands are shown as black dots.}
        \label{fig:hf_egap_and_C_bands}
    \end{figure*}

    We first study the basis set performance for HF ground-state properties, which reflects the quality of the occupied bands.
    The root-mean-square error (RMSE) of the total energy $E_{\tr{tot}}^{\tr{HF}}$, the cohesive energy $E_{\tr{coh}}^{\tr{HF}}$, the equilibrium lattice constant $a_{0}^{\tr{HF}}$, and the equilibrium bulk modulus $B_{0}^{\tr{HF}}$ are presented in \cref{fig:hf_totcoha0b0} for different Gaussian basis sets, where the error is computed with respect to our PW results.
    The errors for each material are shown in Figs.~S7--S10.

    The most obvious trend in \cref{fig:hf_totcoha0b0} is the monotonic decrease of the error of all four properties by following the hierarchy:
    GTH-DZVP, GTH-cc-pVDZ, GTH-cc-pV(T)DZ, GTH-cc-pV(Q)DZ, GTH-cc-pVTZ, GTH-cc-pV(Q)TZ, GTH-cc-pVQZ.
    This ranking is consistent with the flexibility of the basis sets except for GTH-DZVP, which differs from GTH-cc-pVDZ only in the size of the valence basis and the basis parameters.
    The larger error of the GTH-DZVP basis mainly comes from the two beryllium compounds where the contraction coefficients for the valence orbitals of Be are poor and the solid neon where the valence basis is too small (Figs.~S7--S10).
    We verified that a modified GTH-DZVP basis for Be with the contraction coefficients re-computed using our code gives very similar accuracy as our GTH-cc-pVDZ basis.
    Similar reparametrization for the GTH-TZVP and GTH-QZVP basis sets can be performed but is not useful in practice due to the high linear dependencies of these basis sets.

    Among our correlation-consistent basis sets, increasing the zeta-level is significantly more effective at reducing the errors than de-contracting the valence basis, which suggests reasonable transferability of the valence contraction coefficients determined from the atomic HF orbitals, at least for describing the occupied bands (see also the discussion on the HF band structure in \cref{subsec:lowe_vir_bands}).
    The relatively large RMSE in the HF total energy [\cref{fig:hf_totcoha0b0}(a)] (about 4~m\Ha{} even for the GTH-cc-pVQZ basis) is dominated by the two small-core magnesium compounds (Fig.~S7), which inherit the large HF energy error of the magnesium atom, as shown in \cref{tab:valence_basis}.
    Nonetheless, the error in the total energy does not affect computed properties, suggesting a robust and systematic error cancellation in these basis sets.

    \subsection{Low-energy virtual bands: HF band gap and band structure}
    \label{subsec:lowe_vir_bands}

    The RMSEs of the HF band gap calculated using different Gaussian basis sets are summarized in \cref{fig:hf_egap_and_C_bands}(a) (the errors for each material are shown in Fig.~S11), which exhibits an overall trend similar to that discussed in \cref{subsec:occ_bands} for the HF ground-state properties.
    However, a major distinction in the band gap calculations is the significant reduction of error by de-contracting the valence basis of GTH-cc-pVDZ, which indicates limited transferability of the valence contraction coefficients for describing the virtual bands.
    Nonetheless, even without the valence de-contraction, the smallest GTH-cc-pVDZ basis already achieves a RMSE below $0.1$ eV, which is sufficient for most band gap calculations.
    We emphasize that the additional diffuse functions determined in the way described in \cref{subsubsec:gth_diffuse} are crucial for obtaining accurate band gaps for the noble gas solids (Fig.~S11), reducing the error from a few eVs to less than $0.1$ eV in the most extreme case.

    As an example of the performance for the band structure, in \cref{fig:hf_egap_and_C_bands}(b-e) we compare the the valence and low-energy conduction bands of diamond calculated using our Gaussian basis sets to those usings PWs.
    Similar plots for all other materials are displayed in Figs.~S14--S41.
    The smallest GTH-cc-pVDZ basis already gives an accurate description of the valence and the first few conduction bands as shown in \cref{fig:hf_egap_and_C_bands}(b), which is consistent with the good performance observed for the HF ground-state properties (\cref{fig:hf_totcoha0b0}) and the band gap [\cref{fig:hf_egap_and_C_bands}(a)].
    The fixed contraction coefficients in the valence basis of GTH-cc-pVDZ are responsible for the deviations from the PW band structure immediately beyond the first few virtual bands (e.g., at the $\Gamma$ and L points, about $30$ eV above the valence band maximum).
    Using the valence-uncontracted GTH-cc-pV(T)DZ basis fixes this problem and shows quantitative agreement with the PW bands up to about 40~eV, as shown in \cref{fig:hf_egap_and_C_bands}(c).
    The quantitative agreement between the Gaussian and the PW band structures extends to even higher energies by using the GTH-cc-pVTZ ($\sim 70$ eV) and the GTH-cc-pVQZ ($\sim 90$ eV) basis sets, as shown in \cref{fig:hf_egap_and_C_bands}(d) and (e), respectively.

    We note that polarization functions of angular momentum $f$ or higher, which are commonly absent from Gaussian basis sets meant for use in DFT calculations, can be important for the low-energy band structure.
    For example, the authors of ref \citenum{Lee21JCP} highlighted a missing state in the band structure of MgO (the fifth conduction band at the $\Gamma$ point) unless a very large QZ basis set is used ($167$ basis functions per MgO unit).
    By contrast, our calculations on the same system (Fig.~S29) suggest that the observed missing state is primarily an $f$-state localized on the oxygen atom and can already be captured accurately using our GTH-cc-pVTZ basis set with only $59$ basis functions per MgO unit.

    \subsection{Convergence to the full virtual space limit: MP2 ground-state properties}
    \label{subsec:fvs_limit}

    The discussion in the previous two sections have focused on the occupied and the low-lying virtual bands.
    In this section, we study the basis set quality in correlated calculations at the MP2 level, which in principle requires an infinite number of virtual bands in order to reach the CBS limit.
    In a PW basis, the CBS limit is approached in a \emph{dense} manner by increasing the number of virtual bands $n_{\tr{vir}}$ being correlated from low to high energy.
    The Gaussian virtual bands follow the dense manifold of the PW bands in the low-energy regime (as discussed in \cref{subsec:lowe_vir_bands}), but become \emph{sparse} at higher energy, effectively skipping some states.
    Ideally, the correlation energy obtained using either basis should show the asymptotic $n_{\tr{vir}}^{-1}$ convergence (\ref{eq:MP2_corr_nvir}) for sufficiently large $n_{\tr{vir}}$, but this may or may not be achievable with the available computational resources.

\begin{figure}[t]
        \centering
        \includegraphics[scale=0.8]{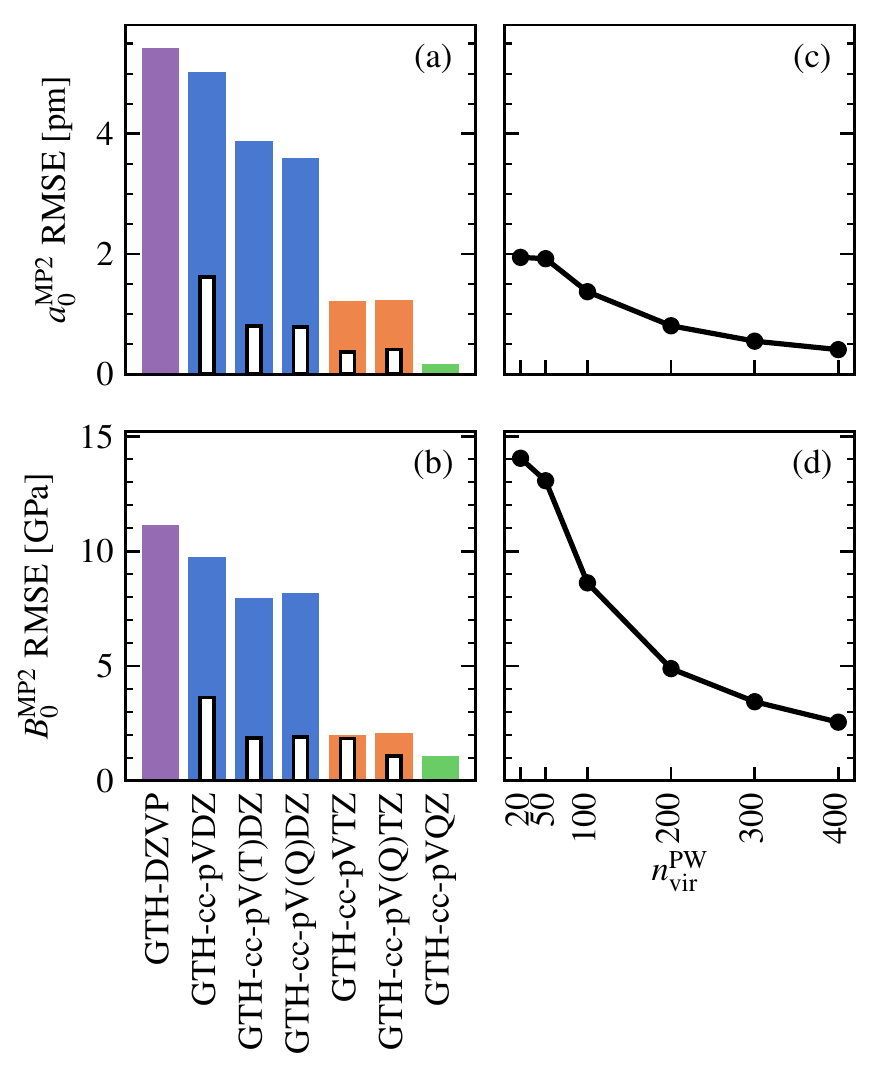}
        \caption{Root-mean-square errors of (a,c) the MP2 equilibrium lattice constant and (b,d) the MP2 equilibrium bulk modulus calculated using (a-b) different Gaussian basis sets and (c-d) the PW basis with increasing number of virtual bands, $n_{\tr{vir}}^{\tr{PW}}$, where the first three points ($n_{\tr{vir}}^{\tr{PW}} = 20$, $50$, and $100$) are chosen to match the size of the virtual space of the GTH-cc-pV$X$Z basis for $X = $ D, T, and Q, respectively.
        In each case, the errors are evaluated against CBS-extrapolated PW results for the seven bulk materials not containing the $s$-block or the noble gas elements: BN, BP, AlN, AlP, C, Si, and SiC.
        For the DZ and TZ Gaussian basis sets (except for GTH-DZVP), the HF-corrected results (see main text for explanation) are shown as the thinner bar with black edge.}
        \label{fig:mp2_a0b0}
    \end{figure}

    For the seven materials that do not contain $s$-block or noble gas elements, i.e.,~BN, BP, AlN, AlP, C, Si, and SiC, reliable extrapolations using \cref{eq:MP2_corr_nvir} can be performed to obtain accurate estimates of $E_{\tr{corr}}^{\tr{MP2}}(\infty)$ in the PW basis (see \cref{sec:computational_details}), from which we compute reference values for the equilibrium lattice constant $a_0^{\tr{MP2}}$ and the equilibrium bulk modulus $B_0^{\tr{MP2}}$.
    The RMSEs of these two properties calculated using different Gaussian basis sets are shown in \cref{fig:mp2_a0b0}(a-b).
    The errors for each material are shown in Figs.~S12 and S13.
    We also compute these two properties using the PW basis without extrapolation for a series of $n_{\tr{vir}}$ and plot the RMSEs in \cref{fig:mp2_a0b0}(c-d).
    The first three points in \cref{fig:mp2_a0b0}(c-d) with $n_{\tr{vir}} = 20$, $50$, and $100$ are chosen to match roughly the number of virtual bands in the GTH-cc-pV$X$Z basis set for $X = $ D, T, and Q, respectively.

    For both properties, the Gaussian basis exhibits the familiar hierarchy observed in the HF band gap calculations (\cref{subsec:lowe_vir_bands}), where increasing the zeta-level significantly improves the accuracy, and de-contracting the valence basis is also effective at the DZ level [\cref{fig:mp2_a0b0}(a-b)].
    The difference between the GTH-DZVP basis and the GTH-cc-pVDZ basis is somewhat smaller than in the HF calculations mainly because the problematic beryllium compounds are not included in the statistics here.

    For correlated calculations with a Gaussian basis, basis set errors enter through both the HF energy and the correlation energy.
    In contrast, for correlated calculations with a PW basis, basis set errors enter through the correlation energy only, because the HF energy is essentially converged with respect to the number of PWs.
    Indeed, the errors observed in \cref{fig:mp2_a0b0}(a-b) for the DZ and TZ bases are dominated by errors in the HF energy, and the results can be significantly improved
    by combining the MP2 correlation energies calculated in a given basis set with the more accurate HF energies obtained from the GTH-cc-pVQZ basis, as shown by the thinner white bars in \cref{fig:mp2_a0b0}(a-b).
    In a similar spirit, one could perform a HF calculation in a large Gaussian basis and then perform an MP2 calculation with some number of frozen virtual orbitals. \cite{Lange20MP,Wang20JCTC}
    Similar corrections are impossible with the original GTH basis set series due to the high linear dependencies at the QZ (or even TZ) level [\cref{fig:condnum_cmp}(a)].

    \begin{figure}[t]
        \centering
        \includegraphics[scale=0.8]{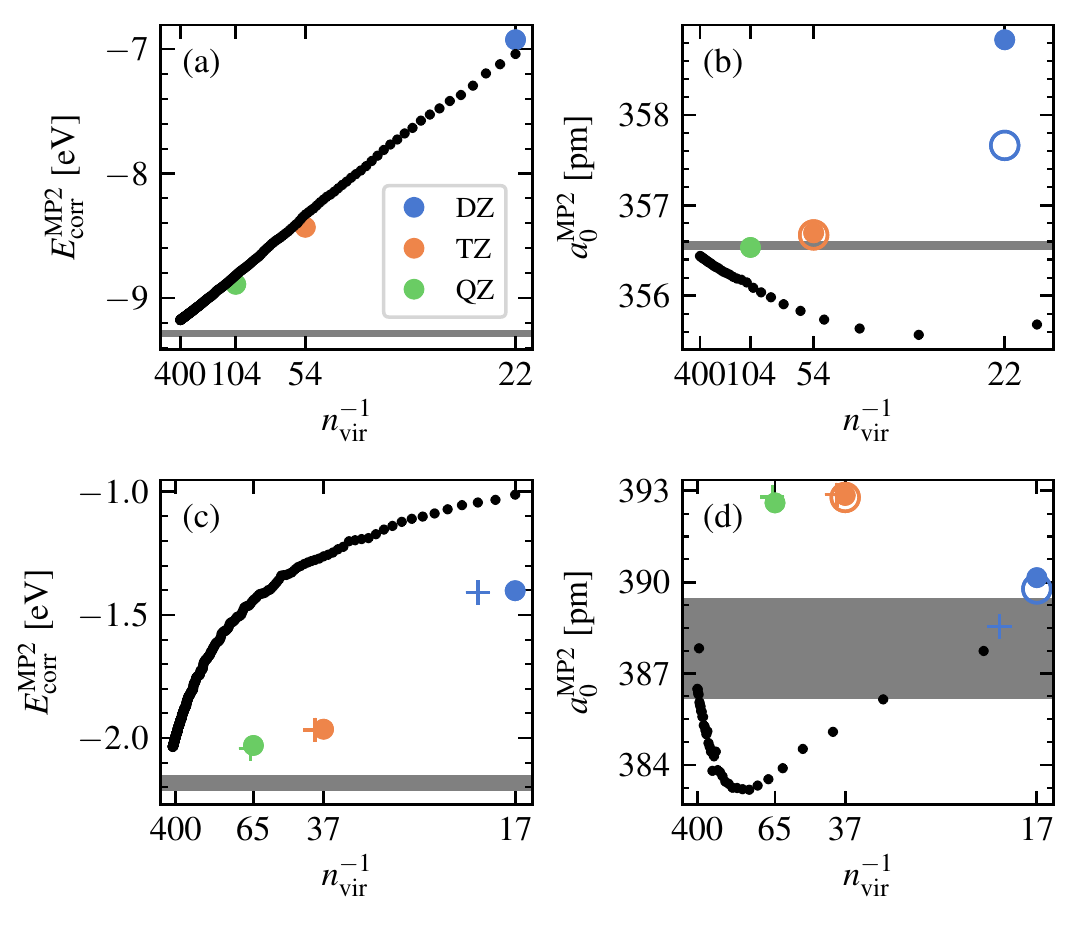}
        \caption{Convergence of the MP2 correlation energy (left column) and the MP2 equilibrium lattice constant (right column) with the number of virtual bands included in the calculations for two materials: diamond (a-b) and LiH (c-d).
        The small-core pseudopotential is used for Li.
        In each case, the small black dots are results obtained using a PW basis, while the blue, orange, and green filled circles are results obtained using the GTH-cc-pVDZ, TZ, and QZ basis sets, respectively.
        For DZ and TZ, the open circles are $a_0^{\tr{MP2}}$ calculated by combining the MP2 correlation energy with the HF energy computed using the QZ basis.
        Gray shaded areas indicate the extrapolated PW results and their uncertainty.
        For LiH, the ``+'' symbols are results obtained by using the PW-resolved Gaussian virtual bands derived from the GTH-cc-pV$X$Z basis sets for the virtual space.
        }
        \label{fig:C_LiH_mp2_qzhf}
    \end{figure}

    For the MP2 calculation of $a_0$ in a PW basis [\cref{fig:mp2_a0b0}(c)], the error of the smallest calculations with only $20$ virtual bands is much smaller than the error of the calculations with the GTH-cc-pVDZ basis without the HF correction, but is similar to the error after the HF correction.
    The situation is different for $B_0$ [\cref{fig:mp2_a0b0}(d)], where the error of the PW calculations with $20$ virtual bands is notably larger than that of the calculations with the GTH-cc-pVDZ basis, even without the HF correction.
    As $n_{\tr{vir}}$ increases, the errors of both properties decay very slowly in the PW basis, especially for small $n_{\tr{vir}}$.
    The largest PW calculations with $n_{\tr{vir}} = 400$ only achieve a RMSE comparable to the HF-corrected GTH-cc-pVTZ basis for $a_0$ and the HF-corrected GTH-cc-pVDZ basis for $B_0$, where the two Gaussian basis sets use only about $50$ (TZ) and $20$ (DZ) virtual bands in the correlated calculations.

    The slow convergence of $a_0$ and $B_0$ with the number of PWs is caused by the small imbalance of the correlation energies evaluated using a fixed $n_{\tr{vir}}$ at the different cell volumes needed for the equation of state.
    This is illustrated in \cref{fig:C_LiH_mp2_qzhf}(a-b) for diamond. At a given lattice constant (here, the equilibrium lattice constant), the MP2 correlation energies evaluated in both the PW and the Gaussian basis sets exhibit the desired $n_{\tr{vir}}^{-1}$ convergence (\ref{eq:MP2_corr_nvir}) [\cref{fig:C_LiH_mp2_qzhf}(a)].
    But only the Gaussian basis sets converge quickly to the CBS limit for $a_0$ [\cref{fig:C_LiH_mp2_qzhf}(b)].
    This behavior occurs because the PW basis is ignorant of the underlying atomic structure and thus exhibits an unphysical sensitivity to the cell volume.
    The situation is even worse in correlated calculations of molecular crystals \cite{DelBen12JCTC,DelBen13JCTC} and free molecules or atoms \cite{Gruneis11JCTC}, due to the large amount of empty space between atoms or molecules.
    The Gaussian basis sets, along with the well-established BSSE correction \cite{Boys70MP,vanDuijneveldt94CR}, are more suitable for describing electron correlation in these systems.

    The poor performance of PWs is exacerbated for elements with hard pseudopotentials, e.g.,~the $s$-block elements with core or semi-core electrons.
    We illustrate this behavior for the ionic crystal LiH, using the small-core pseudopotential for Li.
    As shown in \cref{fig:C_LiH_mp2_qzhf}(c), the convergence of the PW MP2 correlation energy is much slower than for diamond, with the asymptotic $n_{\tr{vir}}^{-1}$ convergence (\ref{eq:MP2_corr_nvir}) not achieved even when $n_{\tr{vir}} \approx 400$.
    This slow convergence yields a large uncertainty in the extrapolated $E_{\tr{corr}}^{\tr{MP2}}(\infty)$ (grey shaded area), which is similar to the GTH-cc-pVQZ result (green circle) obtained with only \rvv{$65$} virtual bands.
    Extrapolation of the Gaussian basis results suggests that the extrapolated PW result is likely an underestimate.
    The convergence in the PW basis is even slower for $a_0$ as shown in \cref{fig:C_LiH_mp2_qzhf}(d), which yields an extrapolated value that is about $4$~pm below that obtained with our Gaussian basis sets.
    The Gaussian basis results converge much more quickly and their correctness is verified by comparison to literature values \cite{Gruneis10JCP} as shown in \cref{tab:LiH_cf_PAW}.

    We end the discussion by showing that the convergence of PW-based correlated calculations can be significantly improved by leveraging a good Gaussian basis, such as the one developed in this work.
    Specifically, we compute the virtual bands in a PW-resolved Gaussian basis \cite{Booth16JCP,Morales20JCP} generated by projecting out the converged PW occupied bands from our GTH-cc-pV$X$Z basis sets, followed by orthonormalization.
    As shown in \cref{fig:C_LiH_mp2_qzhf}(c-d) for LiH, the MP2 calculations that use the PW occupied bands plus the PW-resolved Gaussian virtual bands show significantly faster convergence than those that use the bare PW virtual bands.
    This example shows that our Gaussian basis sets are also useful for PW-based correlated calculations.

    \begin{table}[!t]
        \centering

        \caption{Comparison of the bulk properties of LiH (the small-core pseudopotential is used for Li) to those from Ref.~\citenum{Gruneis10JCP} obtained using PWs and the projector augmented-wave (PAW) method \cite{Blochl94PRB}.
        The Gaussian basis calculations use an unshifted $5\times5\times5$ $k$-point mesh without extrapolation.}
        \label{tab:LiH_cf_PAW}


        \begin{ruledtabular}
        \begin{tabular}{lcccccc}
            & \multicolumn{2}{c}{$a_0$ [pm]} &
              \multicolumn{2}{c}{$B_0$ [GPa]} &
              \multicolumn{2}{c}{$E_{\tr{coh}}$ [eV]} \\
            \cmidrule(lr){2-3}\cmidrule(lr){4-5}\cmidrule(lr){6-7}
            & HF & MP2 & HF & MP2 & HF & MP2 \\
            \hline
            GTH-cc-pVDZ & $410.4$ & $394.3$ & $32.5$ & $39.4$ & $1.81$ & $2.34$ \\
            GTH-cc-pVTZ & $410.2$ & $396.7$ & $32.5$ & $38.5$ & $1.81$ & $2.37$ \\
            GTH-cc-pVQZ & $410.1$ & $396.5$ & $32.6$ & $38.6$ & $1.81$ & $2.38$ \\
            PAW+PW & $411.1$ & $397.1$ & $32$ & $38$ & $1.79$ & $2.39$
        \end{tabular}
        \end{ruledtabular}
    \end{table}

\section{Conclusion}
\label{sec:conclusion}

    To conclude, we extended Dunning's strategy for constructing correlation-consistent Gaussian basis sets to periodic systems by controlling the size of the valence basis to reach a balance between accuracy and numerical stability.
    The generated GTH-cc-pV$X$Z basis sets are found to be well-conditioned for solid-state calculations and show fast convergence to the CBS limit at both mean-field and correlated levels of theory on a number of bulk properties.
    Our scheme can also be used straightforwardly to design all-electron basis sets for solids, which will differ only by the addition of primitives with large exponents that do not significantly contribute to linear dependencies.

    Although our basis sets were tested using MP2, they will be valuable in work using more accurate ab initio correlated methods, such as coupled-cluster theory \cite{McClain17JCTC,Gruber18PRX,Zhang19FM,Lau21JPCL,Lange21JCP,Callahan21JCP},
    auxiliary field quantum Monte Carlo \cite{Ma15PRL,Rudshteyn20JCTC,Morales20JCP,Malone20JCTC,Shi21JCP},
    or quantum embedding approaches \cite{Bulik14JCP,Rusakov19JCTC,Iskakov20PRB,Yeh21PRB,Yeh21PRB,Pham20JCTC,Cui20JCTC,Zhu20JCTC,Ye20JCTC,Zhu21PRX}.
    In particular, our basis sets remain to be tested on three-dimensional metals, where MP2 is inapplicable.\cite{Shepherd13PRL}
    Reliable and standardized Gaussian basis sets for periodic systems also call for the development of optimized auxiliary basis sets for density fitting \cite{Weigend98CPL,Weigend02PCCP,Weigend08JCC}.
    Finally, future work will proceed down the periodic table to obtain performant Gaussian basis sets for more elements.
    Of special interest are the $d$ and $f$-block metals due to their appearance in a variety of functional materials \cite{Keimer15Nature,Du20Nature,Giuffredi21ACSMAu}, whose accurate description demands correlated electronic structure theories.

\section*{Acknowledgements}

We thank Dr.~Verena Neufeld for helpful discussions.
This work was supported by the National Science Foundation under Grant No.\
OAC-1931321.  We acknowledge computing resources from Columbia University's
Shared Research Computing Facility project, which is supported by NIH Research
Facility Improvement Grant 1G20RR030893-01, and associated funds from the New
York State Empire State Development, Division of Science Technology and
Innovation (NYSTAR) Contract C090171, both awarded April 15, 2010. The Flatiron
Institute is a division of the Simons Foundation.

\section*{Supporting Information}

\rvv{See the supporting information for (i) plot of the condition number of the overlap matrix of commonly used Gaussian basis sets for 16 bulk materials, (ii) convergence of the HF and MP2 bulk properties for diamond with respect to the size of the valence basis of carbon, (iii) comparison of the exponents of the GTH-cc-pV$X$Z basis sets with the all-electron cc-pV$X$Z series for all elements studied in this work, (iv) increment of the atomic CCSD correlation energy as a function of the number of polarization functions for all elements studied in this work, (v) material-specific errors at both HF and MP2 levels for all properties reported in \cref{fig:hf_totcoha0b0,fig:hf_egap_and_C_bands,fig:mp2_a0b0}, (vi) band structure calculated using the GTH-cc-pV$X$Z basis sets compared to the PW CBS result for all materials studied in this work, (vii) detailed basis size information of the GTH-cc-pV$X$Z basis sets, and (viii) lattice structure and experimental lattice constants for all materials studied in this work.}

\section*{Data availability statement}
The data that support the findings of this study are available from the
corresponding author upon reasonable request.

\bibliography{refs}

\raggedbottom

\end{document}


\maketitle

    \tableofcontents

    \vspace{3em}

    \hspace{2em}Note: figures and equations appearing in the main text will be referred as
``Fig.\ Mxxx'' and ``Eq.\ Mxxx'' in this Supplementary Material document.


    \section{Supplementary figures}

    \begin{figure}[h]
        \centering
        \includegraphics[width=0.6\linewidth]{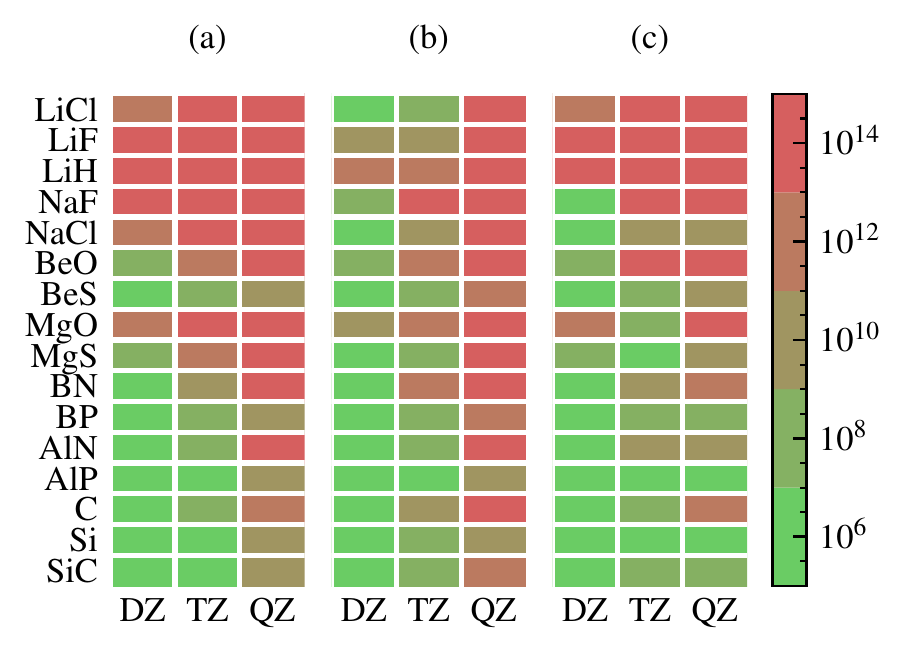}
        \caption{Same plot as Fig.~M1 for the condition numbers of (a) cc-pV$X$Z, (b) def2-SVP (DZ), def2-TZVPP (TZ), and def2-QZVPP (QZ), and (c) ccECP-cc-pV$X$Z. For ccECP, the "reg" basis is used for Li and Be, and the "helium-core" basis is used for Mg to match the number of electrons in the small-core GTH pseudopotentials.}
    \end{figure}

    \begin{figure}[h]
        \centering
        \includegraphics[width=0.5\linewidth]{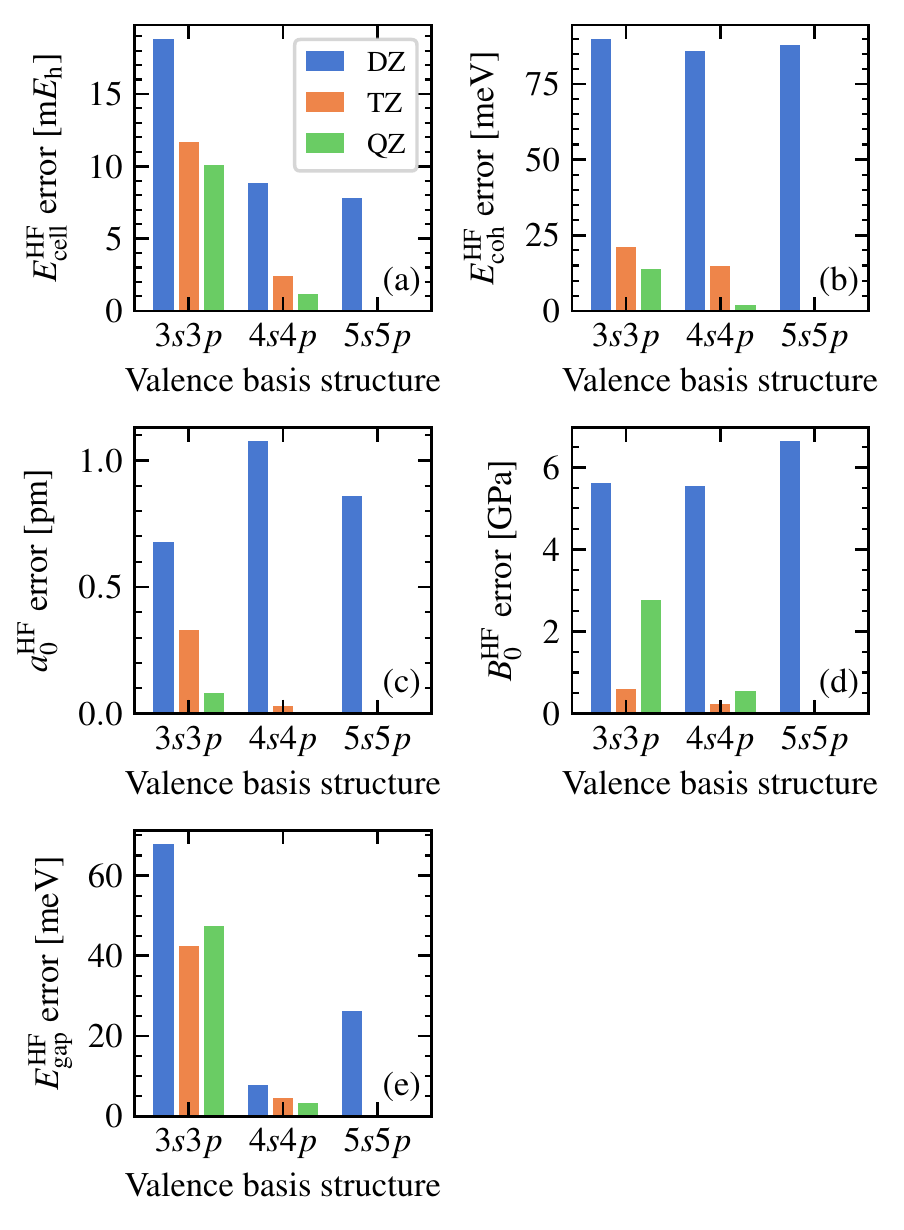}
        \caption{Properties of bulk diamond calculated at HF level using the seven correlation consistent basis sets generated with different valence basis (DZ, TZ, and QZ for $3s3p$ and $4s4p$ and DZ for $5s5p$). (a) Total energy, (b) cohesive energy, (c) equilibrium lattice constant, (d) equilibrium bulk modulus, and (e) band gap.
        Panels (a) and (e) are the same as Fig.~M2(e) and (f), respectively.}
        \label{fig:hf_all}
    \end{figure}

    \begin{figure}[h]
        \centering
        \includegraphics[width=0.5\linewidth]{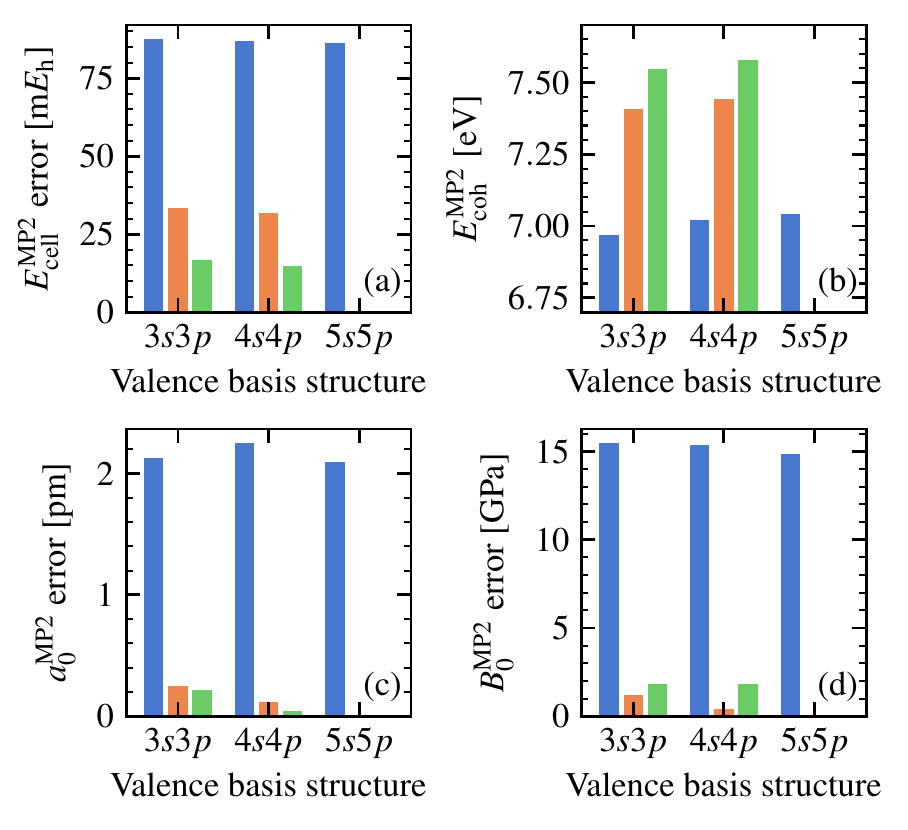}
        \caption{Same plot as \cref{fig:hf_all}(a-d) for MP2 except that the original values instead of the errors are shown for the MP2 cohesive energy in (b).}
        \label{fig:mp2_all}
    \end{figure}

    \begin{figure}[h]
        \centering
        \includegraphics[width=1.0\linewidth]{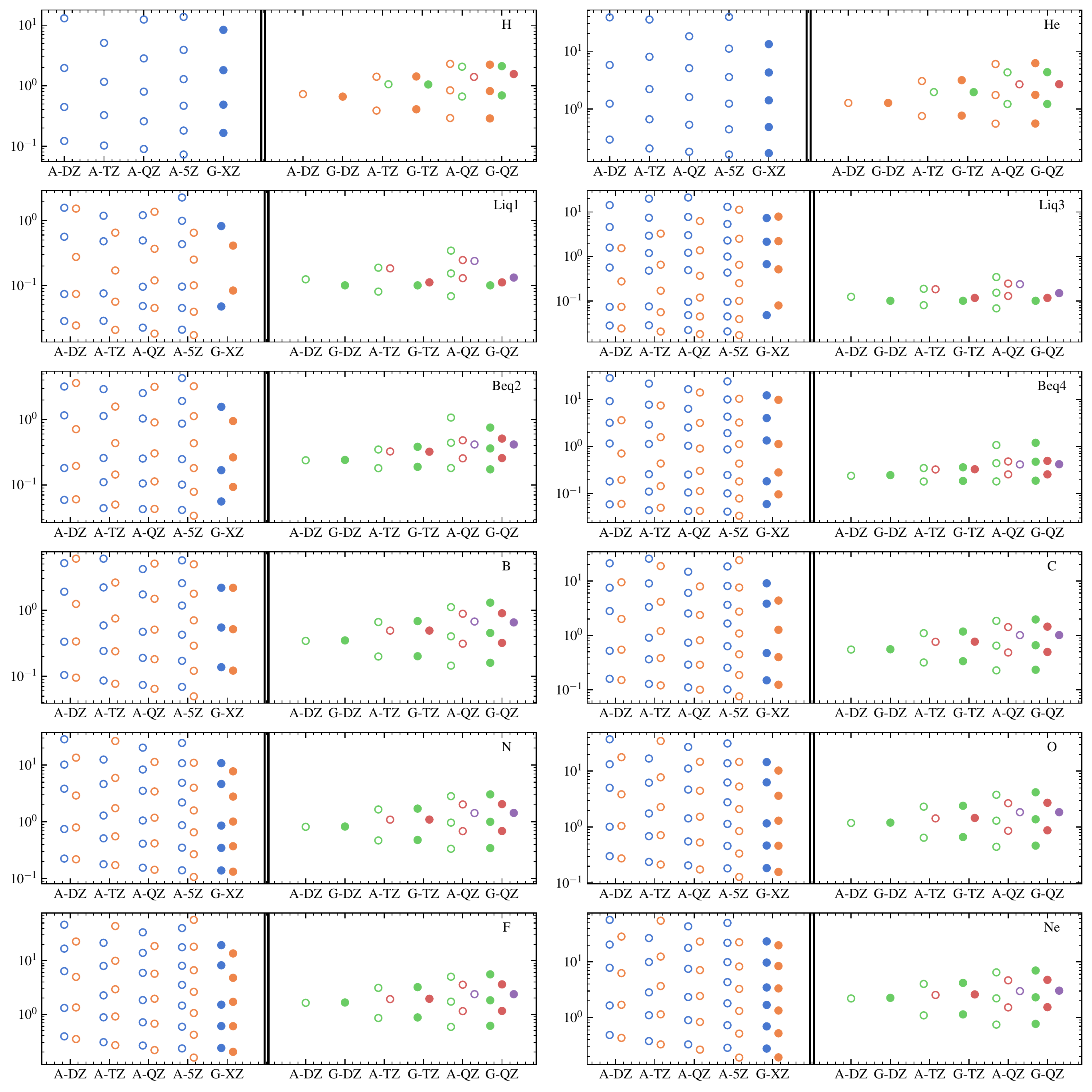}
        \caption{Comparing the exponents in our GTH-cc-pV$X$Z basis sets (labelled as "G-XZ") with those in the all-electron cc-pV$X$Z basis sets (labelled as "A-XZ") for elements from the first two rows.
        The exponents for the valence basis and the polarization functions are separately shown for each element.
        Color scheme: blue, orange, green, red, and violet for $s$, $p$, $d$, $f$, and $g$ angular momentum, respectively.}
        \label{fig:expn_12row}
    \end{figure}

    \begin{figure}[h]
        \centering
        \includegraphics[width=1.0\linewidth]{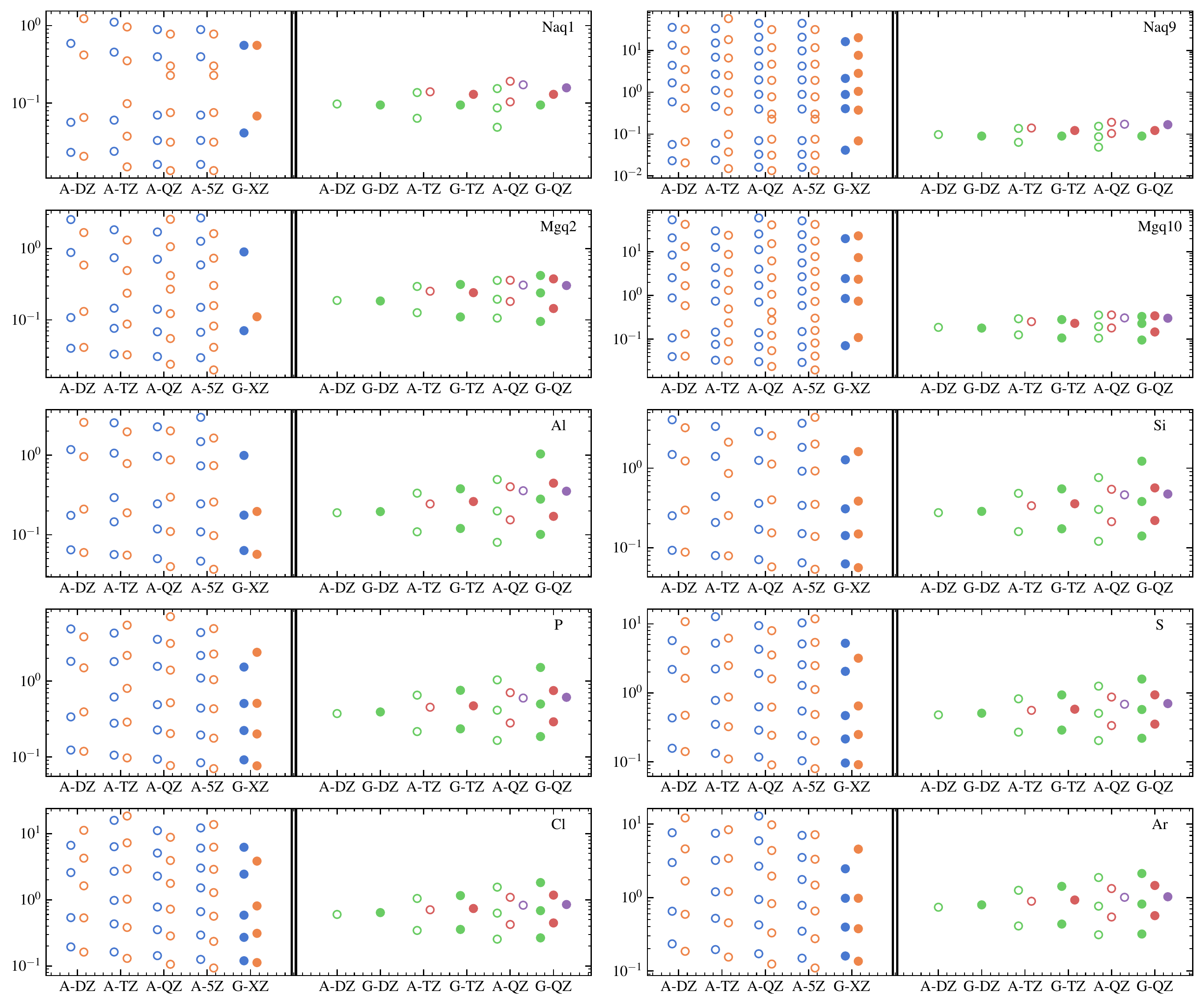}
        \caption{Same plot as \cref{fig:expn_12row} for elements from the third row.}
    \end{figure}

    \begin{figure}[h]
        \centering
        \includegraphics[width=1.0\linewidth]{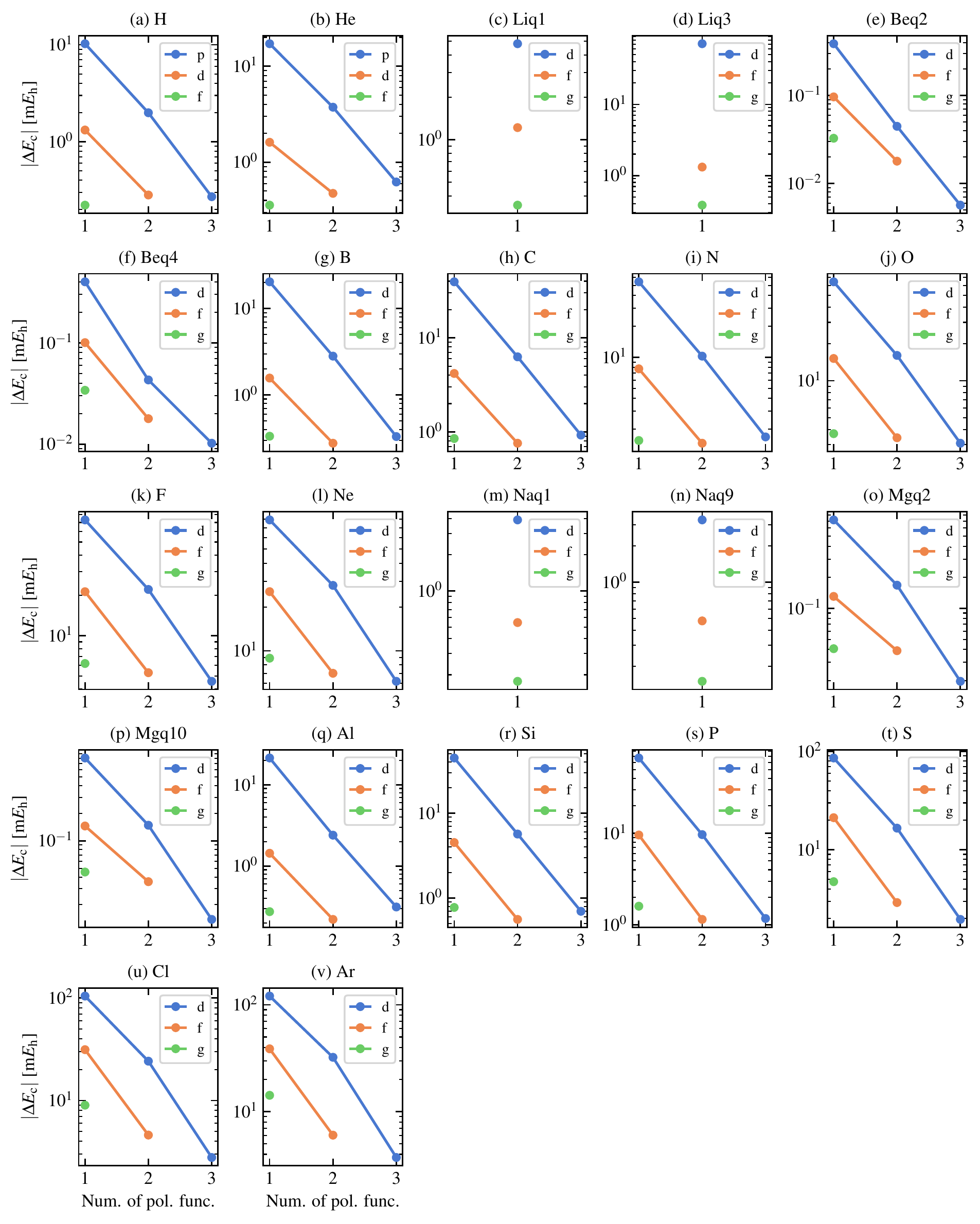}
        \caption{Same plot as Fig.~M2(c) for all elements studied in this work.
        For $s$-block metals, the number after "q" denotes the number of active electrons (i.e.,~not covered by the pseudopotentials).}
    \end{figure}

    \clearpage

    \begin{figure}[p]
        \centering
        \includegraphics[width=1.0\linewidth]{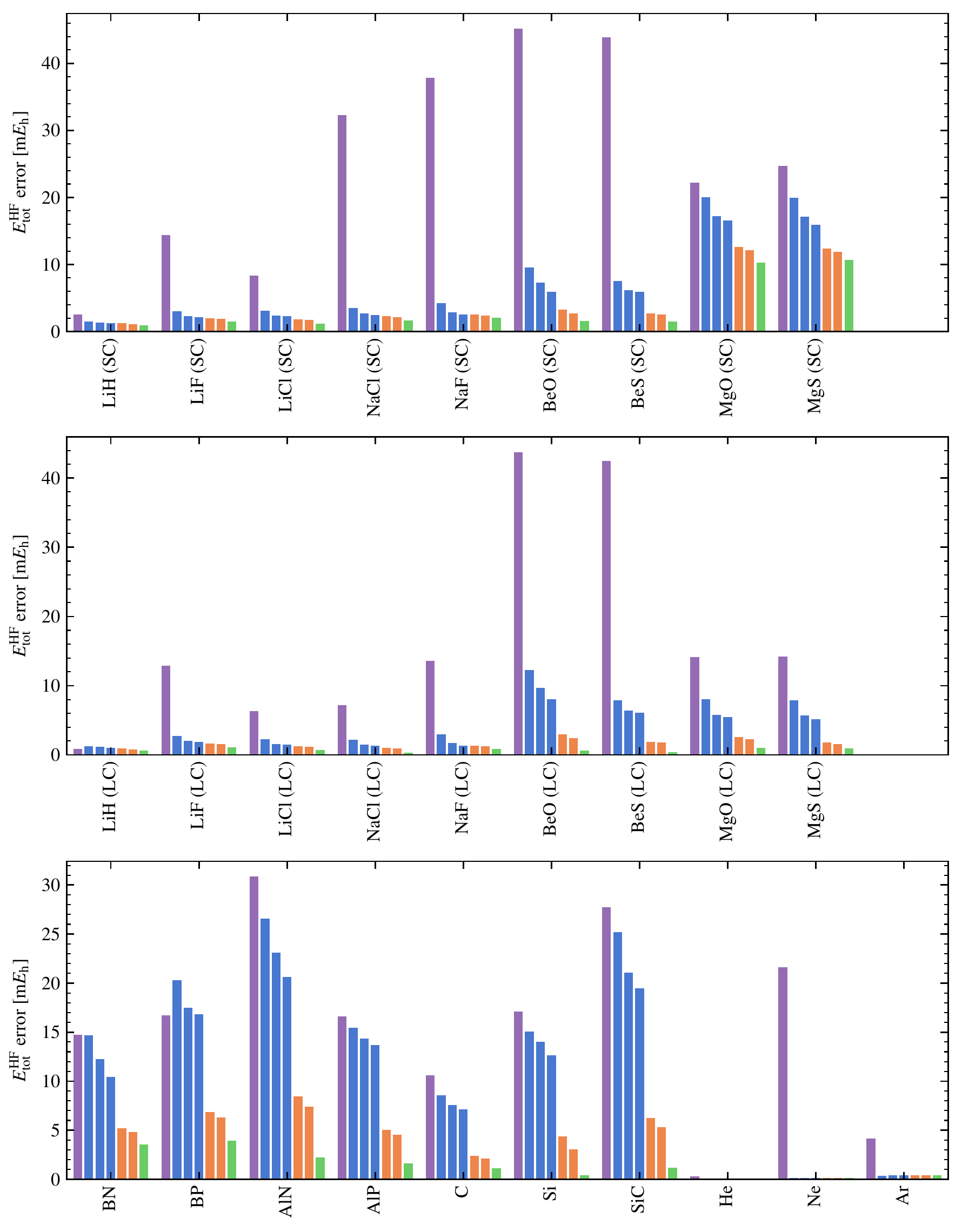}
        \caption{Error of the HF total energy evaluated by different Gaussian basis sets compared to the PW benchmark using a $5\times5\times5$ $k$-point mesh to sample the Brillouin zone.
        For each formula, from left to right: GTH-DZVP (violet), GTH-cc-pVDZ (blue), GTH-cc-pV(T)DZ (blue), GTH-cc-pV(Q)DZ (blue), GTH-cc-pVTZ (orange), GTH-cc-pV(Q)TZ (orange), and GTH-cc-pVQZ (green).}
        \label{fig:hferreach_etot}
    \end{figure}

    \clearpage

    \begin{figure}[p]
        \centering
        \includegraphics[width=1.0\linewidth]{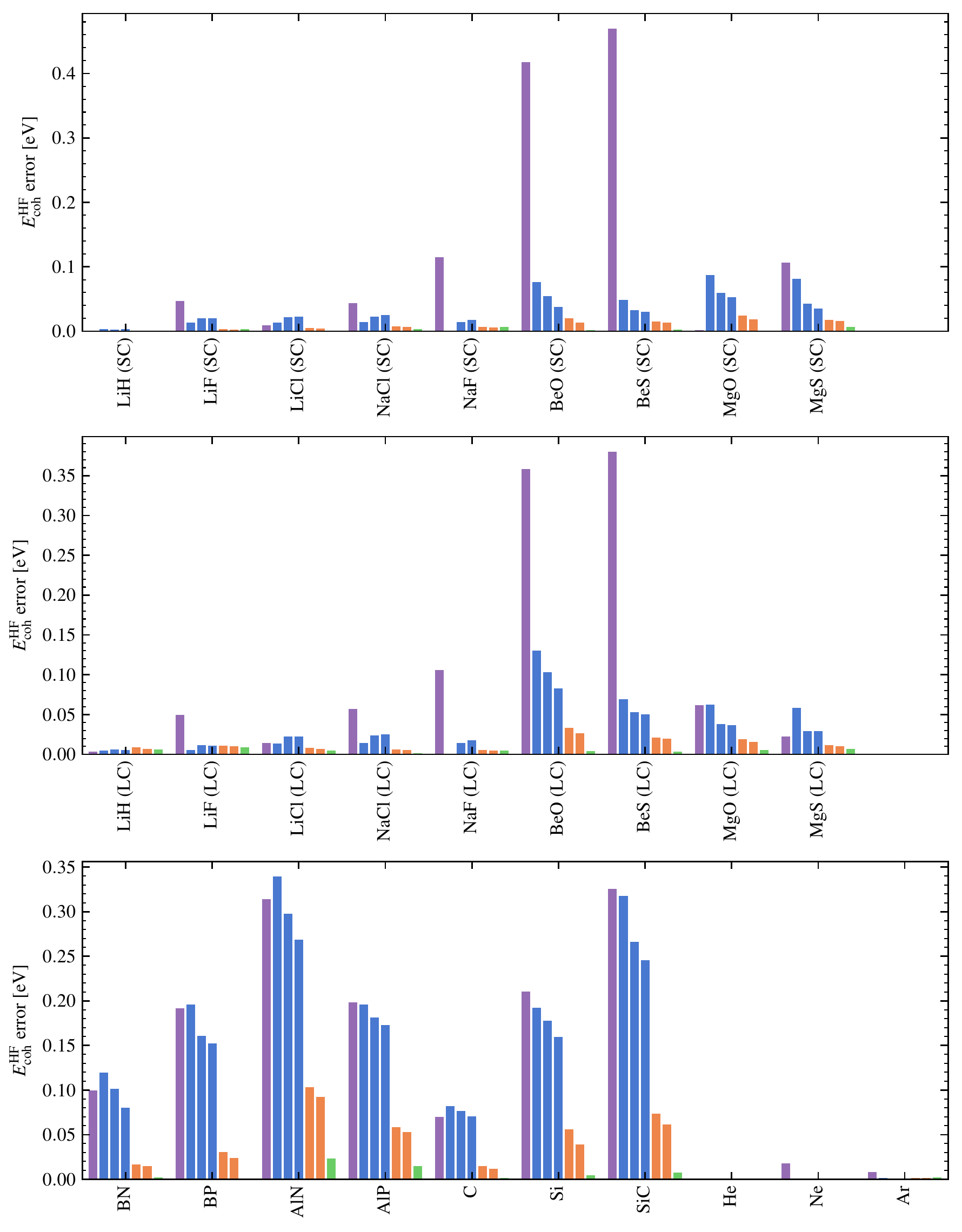}
        \caption{Same as \cref{fig:hferreach_etot} for the HF cohesive energy.}
    \end{figure}

    \clearpage

    \begin{figure}[p]
        \centering
        \includegraphics[width=1.0\linewidth]{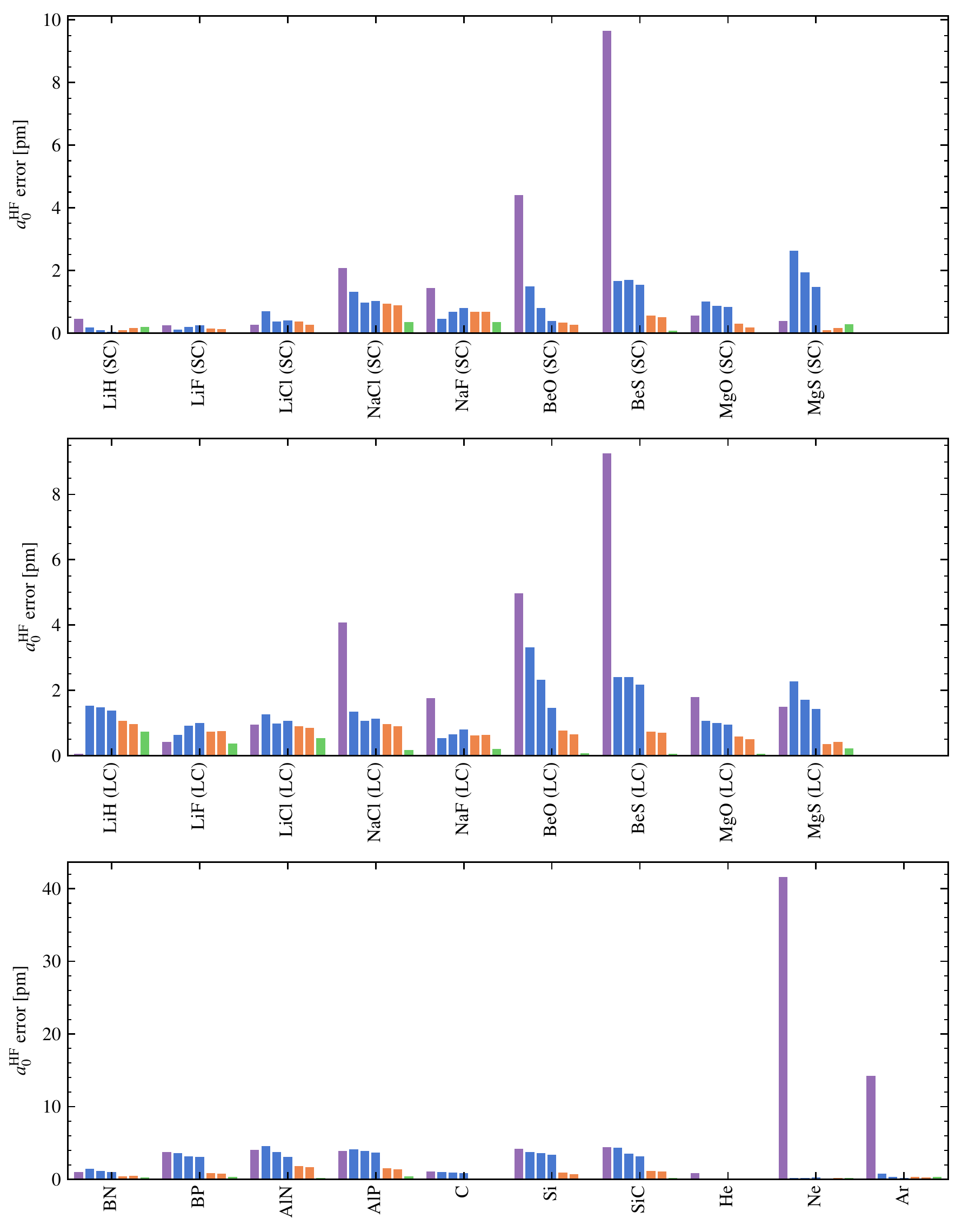}
        \caption{Same as \cref{fig:hferreach_etot} for the HF equilibrium lattice constant evaluated using a $3\times3\times3$ $k$-point mesh.}
    \end{figure}

    \clearpage

    \begin{figure}[p]
        \centering
        \includegraphics[width=1.0\linewidth]{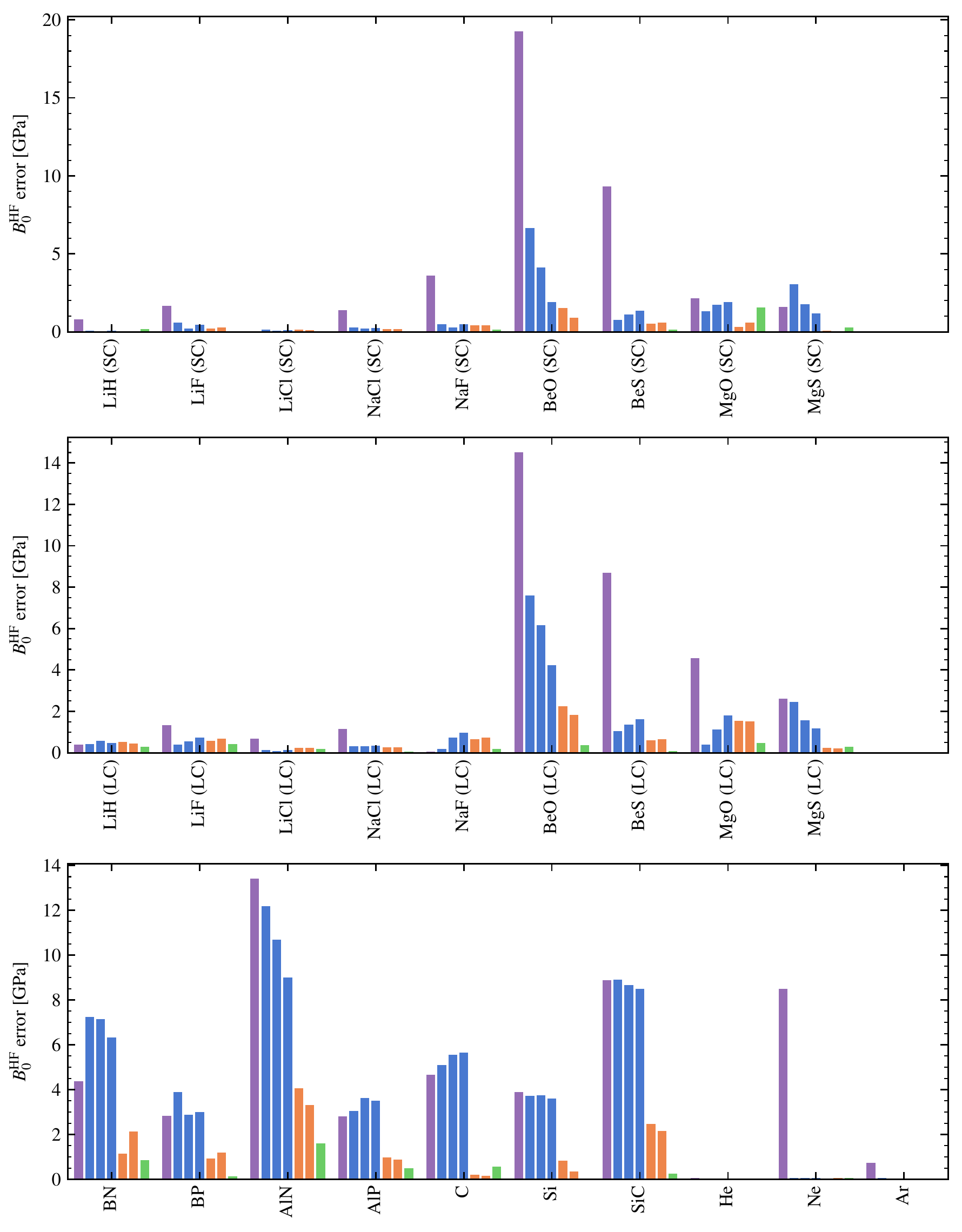}
        \caption{Same as \cref{fig:hferreach_etot} for the HF equilibrium bulk modulus evaluated using a $3\times3\times3$ $k$-point mesh.}
    \end{figure}

    \clearpage

    \begin{figure}[p]
        \centering
        \includegraphics[width=1.0\linewidth]{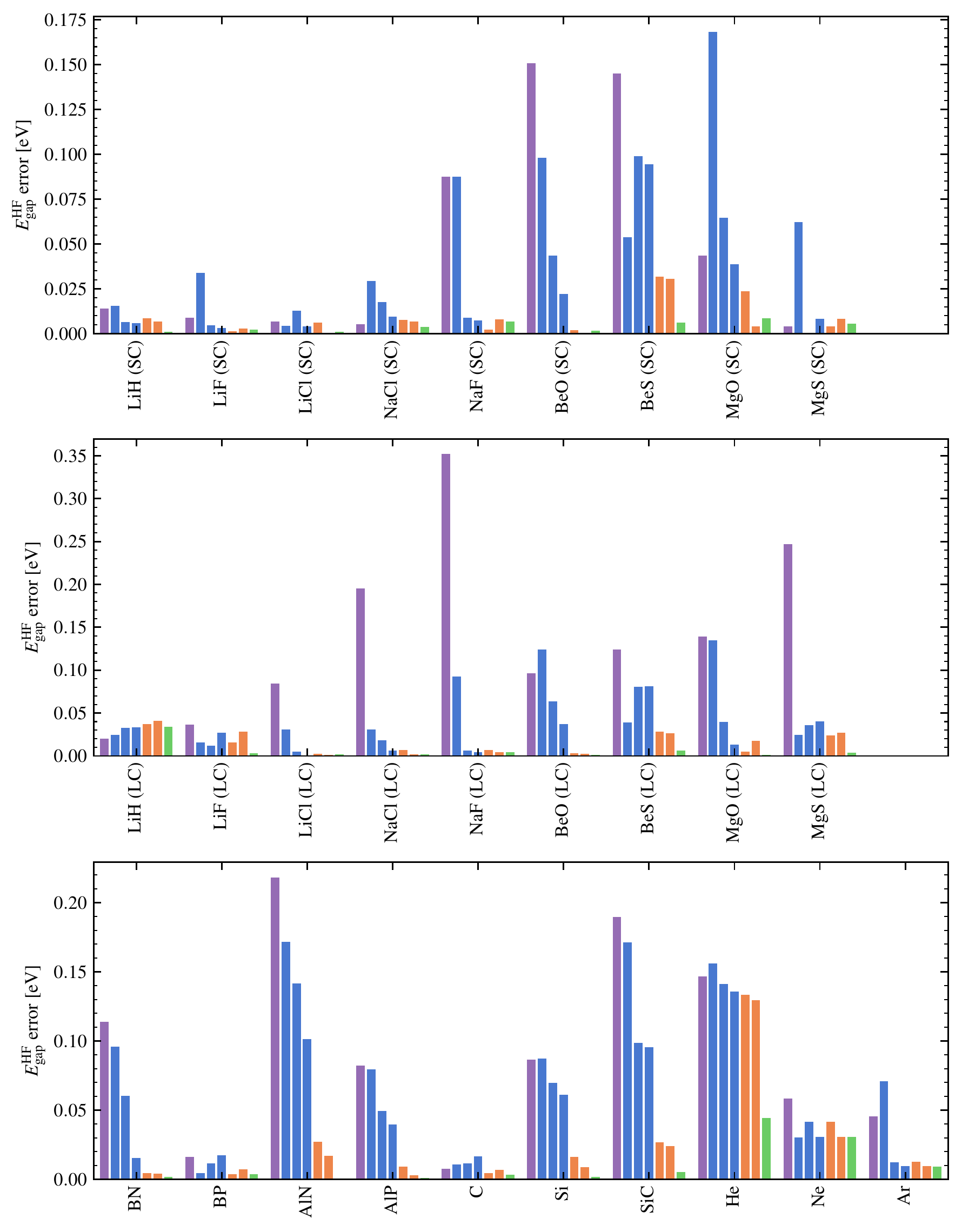}
        \caption{Same as \cref{fig:hferreach_etot} for the HF band gap.}
    \end{figure}

    \clearpage

    \begin{figure}[h]
        \centering
        \includegraphics[width=1.0\linewidth]{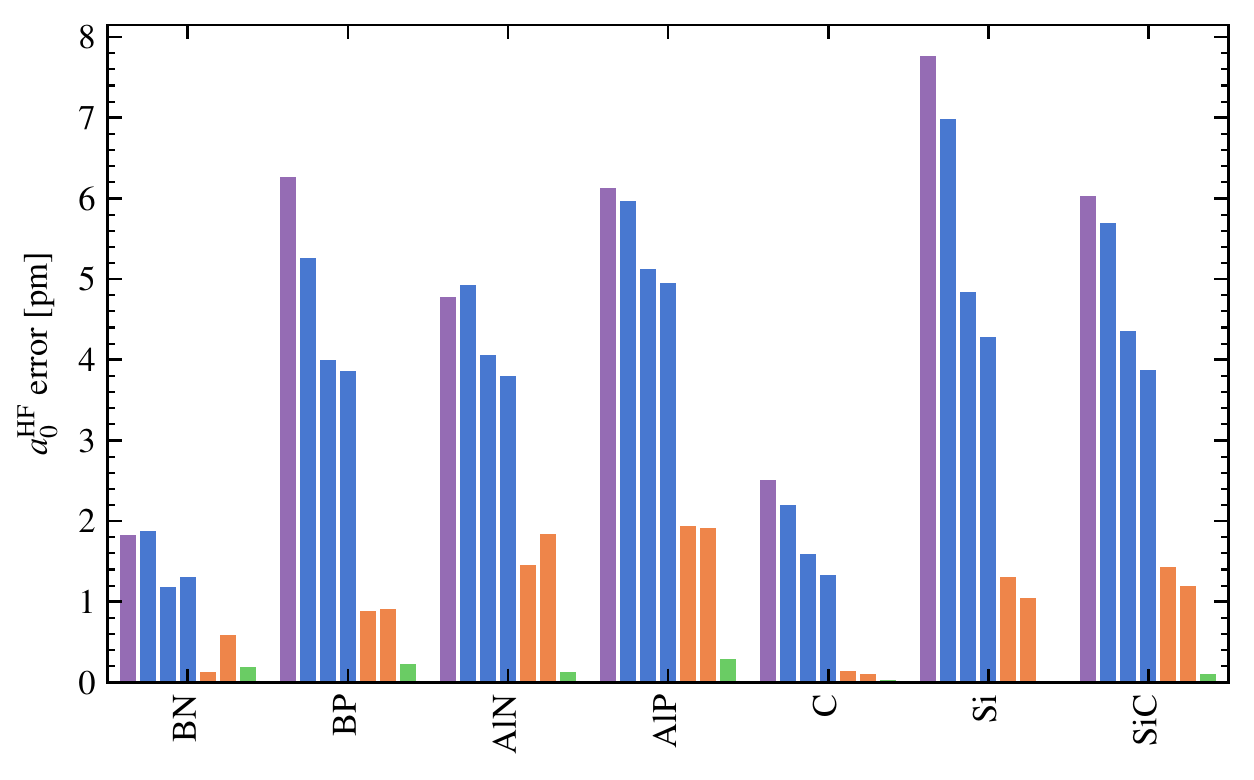}
        \caption{Same as \cref{fig:hferreach_etot} for the MP2 equilibrium lattice constant evaluated using a $3\times3\times3$ $k$-point mesh.}
    \end{figure}

    \begin{figure}[h]
        \centering
        \includegraphics[width=1.0\linewidth]{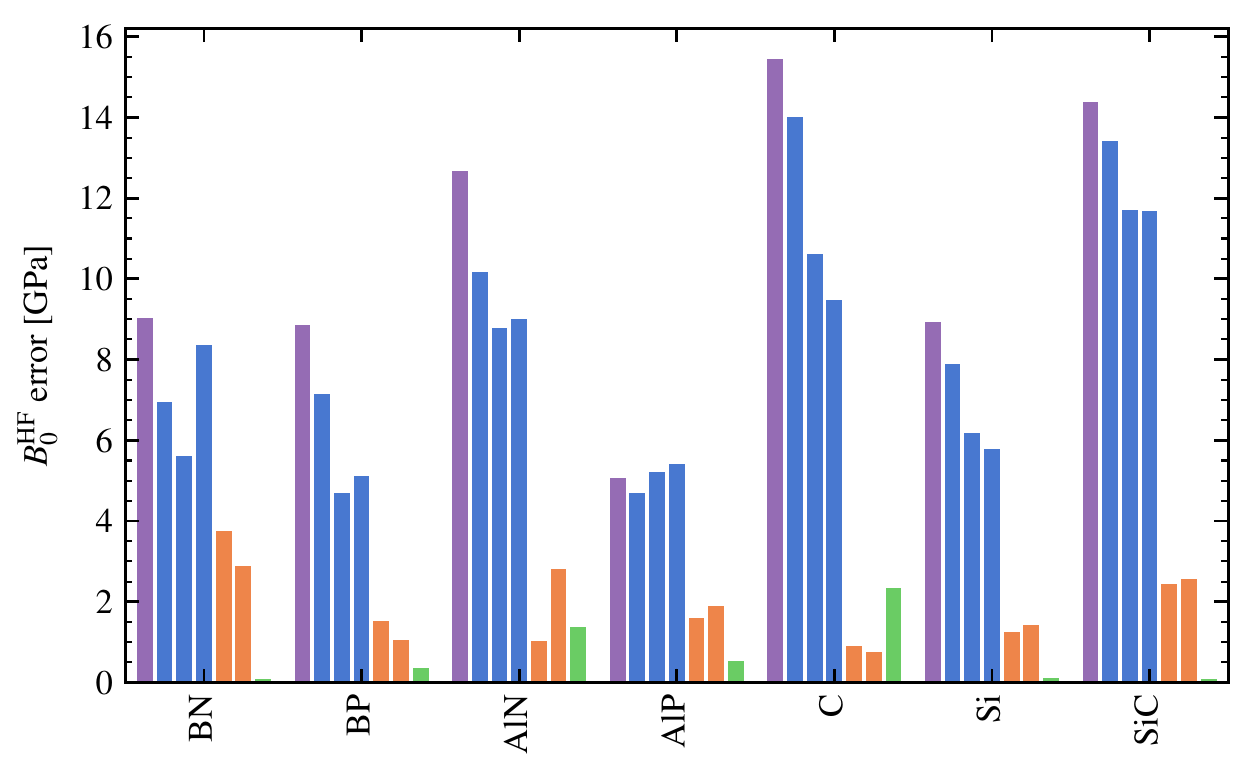}
        \caption{Same as \cref{fig:hferreach_etot} for the MP2 equilibrium bulk modulus evaluated using a $3\times3\times3$ $k$-point mesh.}
    \end{figure}

    \clearpage

    \begin{figure}[h]
    \centering
    \includegraphics[width=1.0\linewidth]{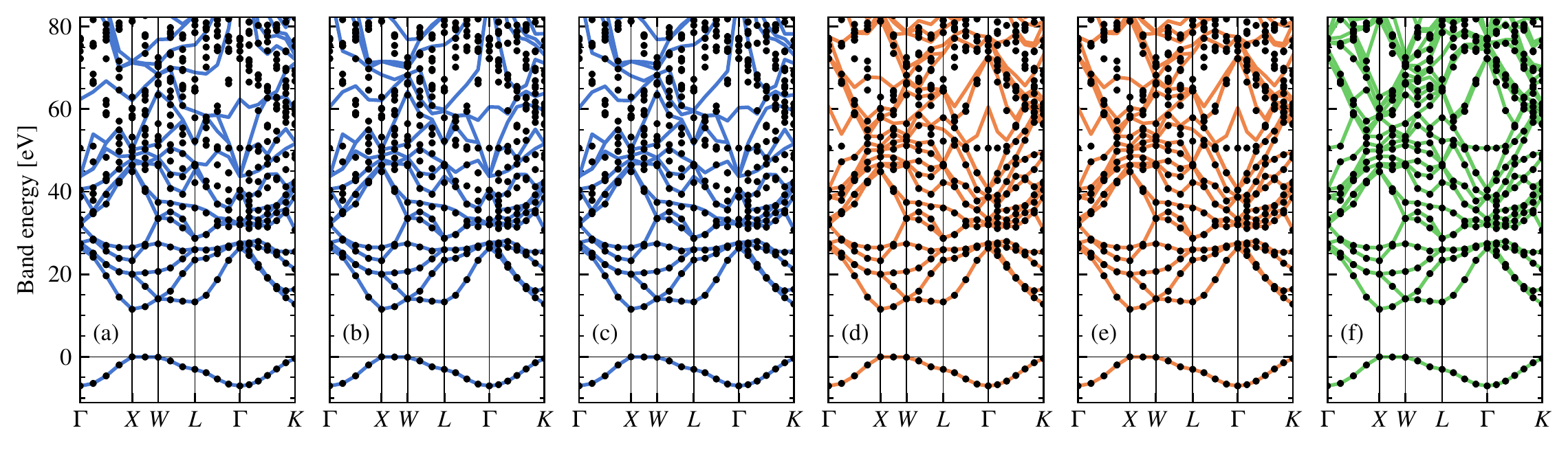}
    \caption{Band structure of LiH (large-core) calculated using different Gaussian basis sets (a) GTH-cc-pVDZ, (b) GTH-cc-pV(T)DZ, (c) GTH-cc-pV(Q)DZ, (d) GTH-cc-pVTZ, (e) GTH-cc-pV(Q)TZ, and (f) GTH-cc-pVQZ, compared to the PW results (black dots in each plot).}
    \label{fig:bands_LiH_large}
\end{figure}

\begin{figure}[h]
    \centering
    \includegraphics[width=1.0\linewidth]{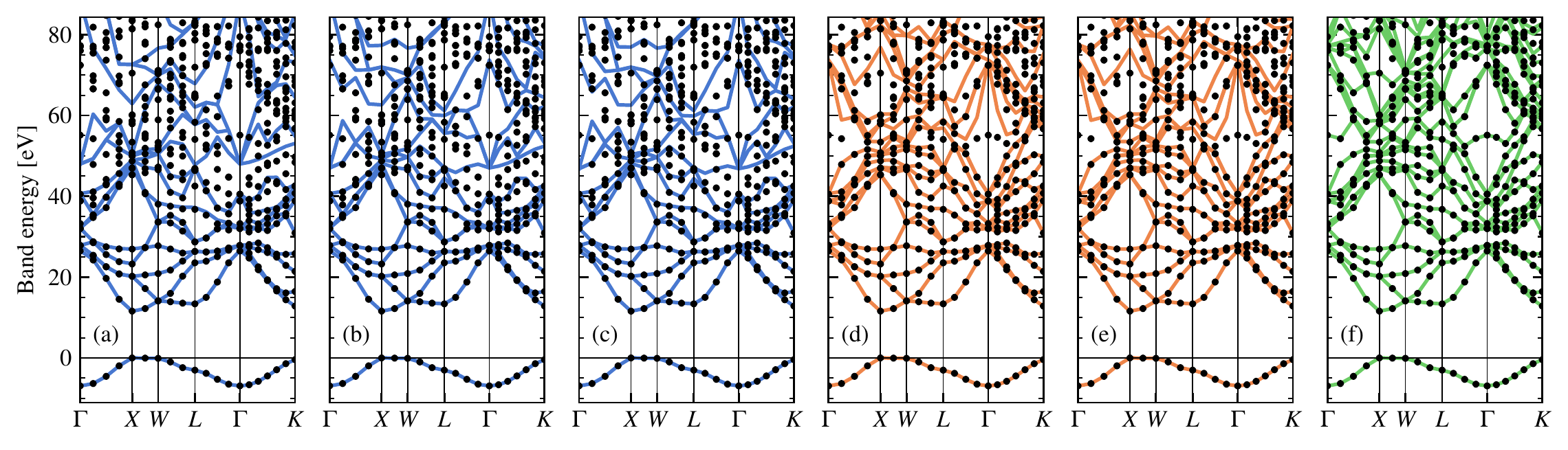}
    \caption{Same plot as \ref{fig:bands_LiH_large} for the band structure of LiH (small-core).}
    \label{fig:bands_LiH_small}
\end{figure}

\begin{figure}[h]
    \centering
    \includegraphics[width=1.0\linewidth]{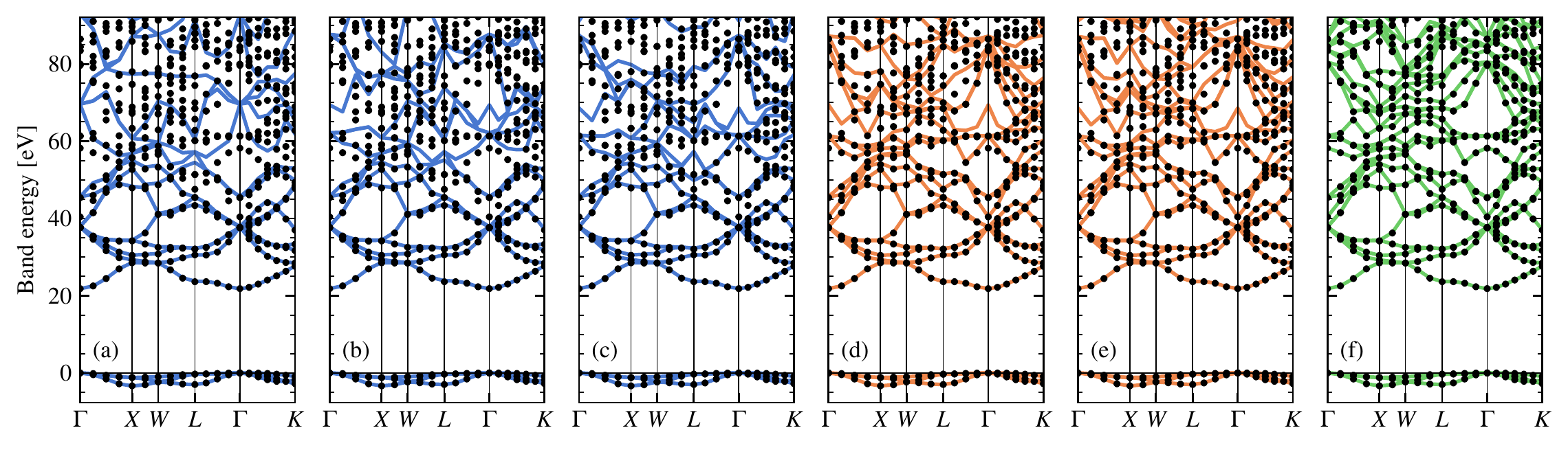}
    \caption{Same plot as \ref{fig:bands_LiH_large} for the band structure of LiF (large-core).}
    \label{fig:bands_LiF_large}
\end{figure}

\begin{figure}[h]
    \centering
    \includegraphics[width=1.0\linewidth]{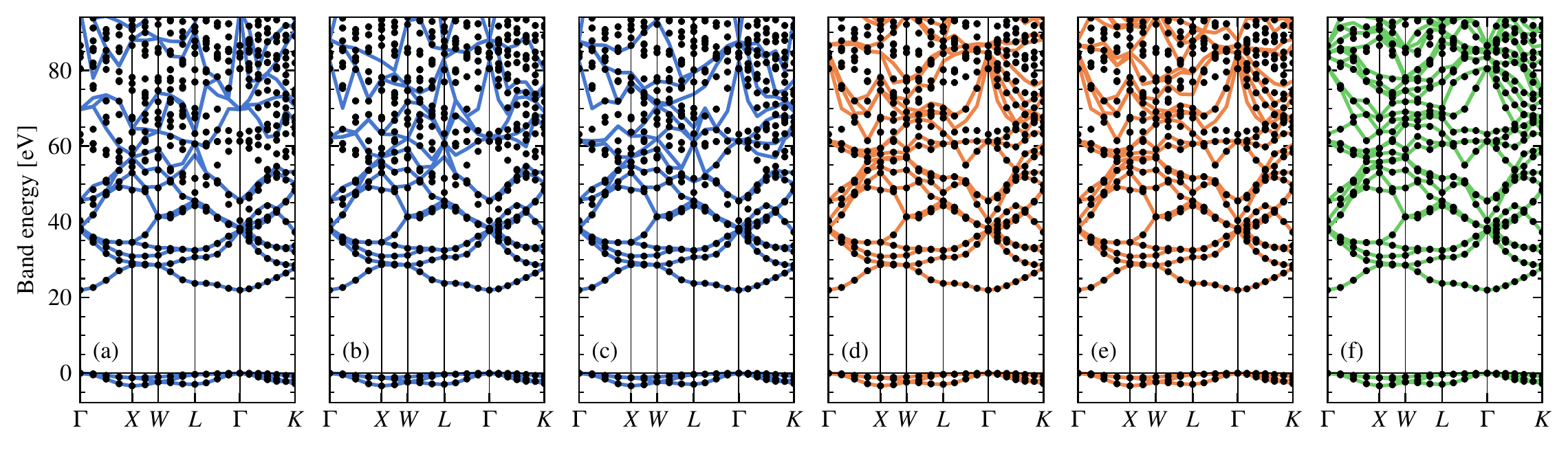}
    \caption{Same plot as \ref{fig:bands_LiH_large} for the band structure of LiF (small-core).}
    \label{fig:bands_LiF_small}
\end{figure}

\begin{figure}[h]
    \centering
    \includegraphics[width=1.0\linewidth]{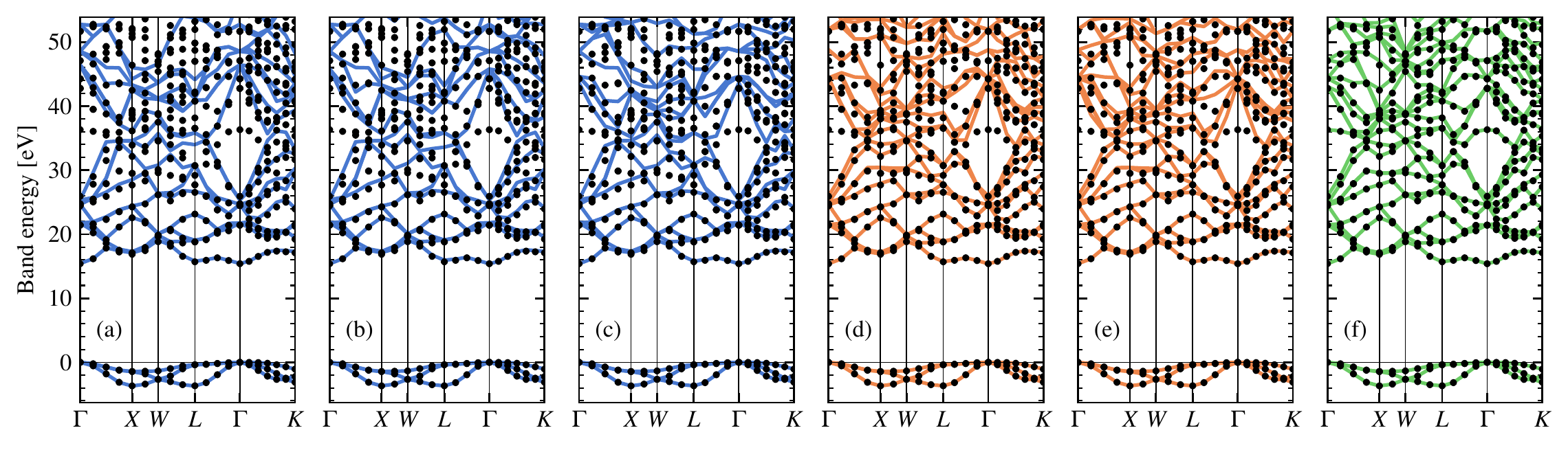}
    \caption{Same plot as \ref{fig:bands_LiH_large} for the band structure of LiCl (large-core).}
    \label{fig:bands_LiCl_large}
\end{figure}

\begin{figure}[h]
    \centering
    \includegraphics[width=1.0\linewidth]{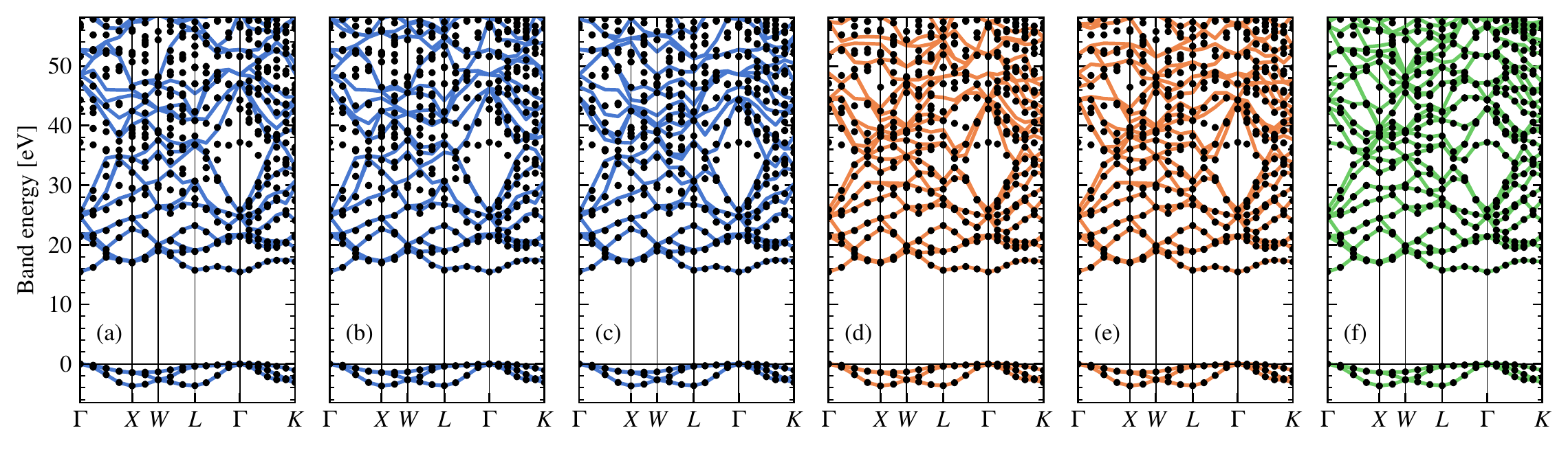}
    \caption{Same plot as \ref{fig:bands_LiH_large} for the band structure of LiCl (small-core).}
    \label{fig:bands_LiCl_small}
\end{figure}

\begin{figure}[h]
    \centering
    \includegraphics[width=1.0\linewidth]{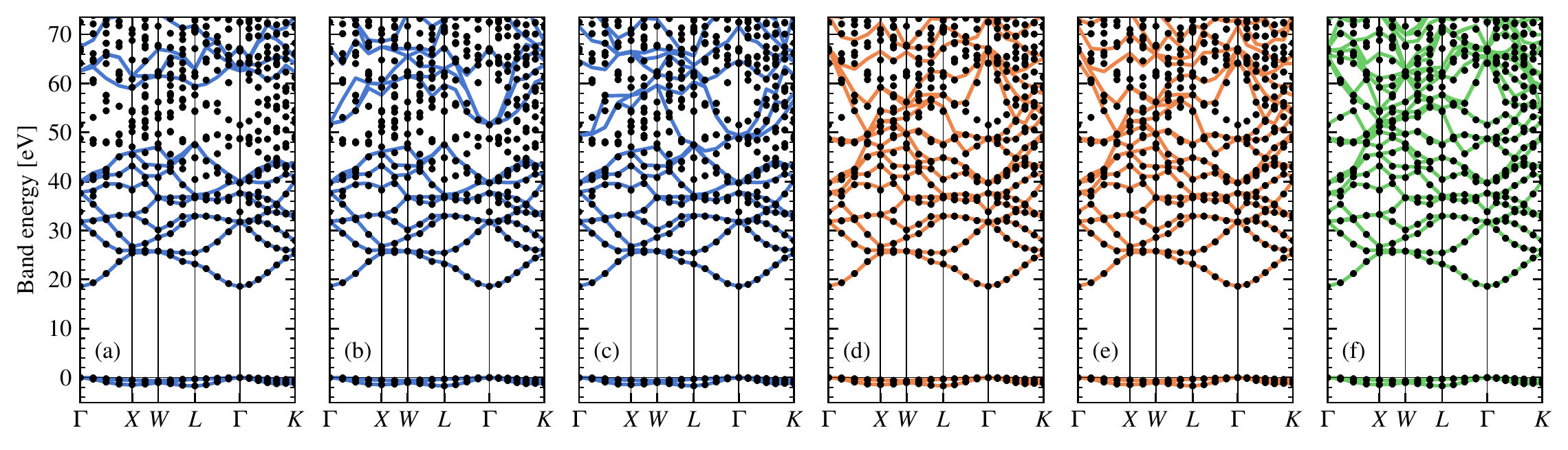}
    \caption{Same plot as \ref{fig:bands_LiH_large} for the band structure of NaF (large-core).}
    \label{fig:bands_NaF_large}
\end{figure}

\begin{figure}[h]
    \centering
    \includegraphics[width=1.0\linewidth]{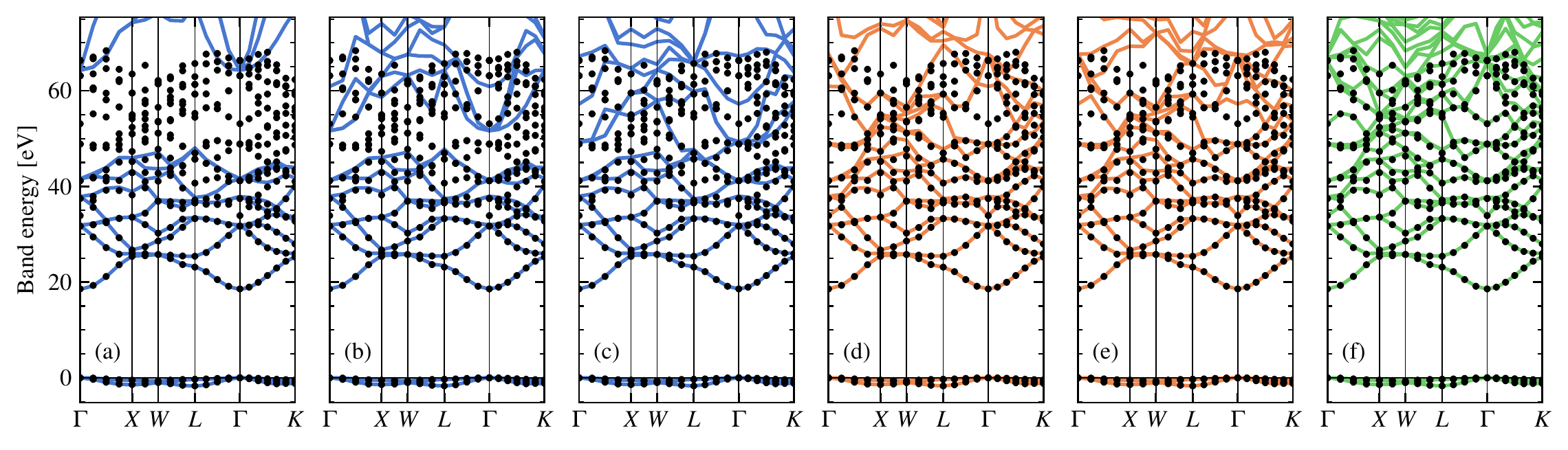}
    \caption{Same plot as \ref{fig:bands_LiH_large} for the band structure of NaF (small-core).}
    \label{fig:bands_NaF_small}
\end{figure}

\begin{figure}[h]
    \centering
    \includegraphics[width=1.0\linewidth]{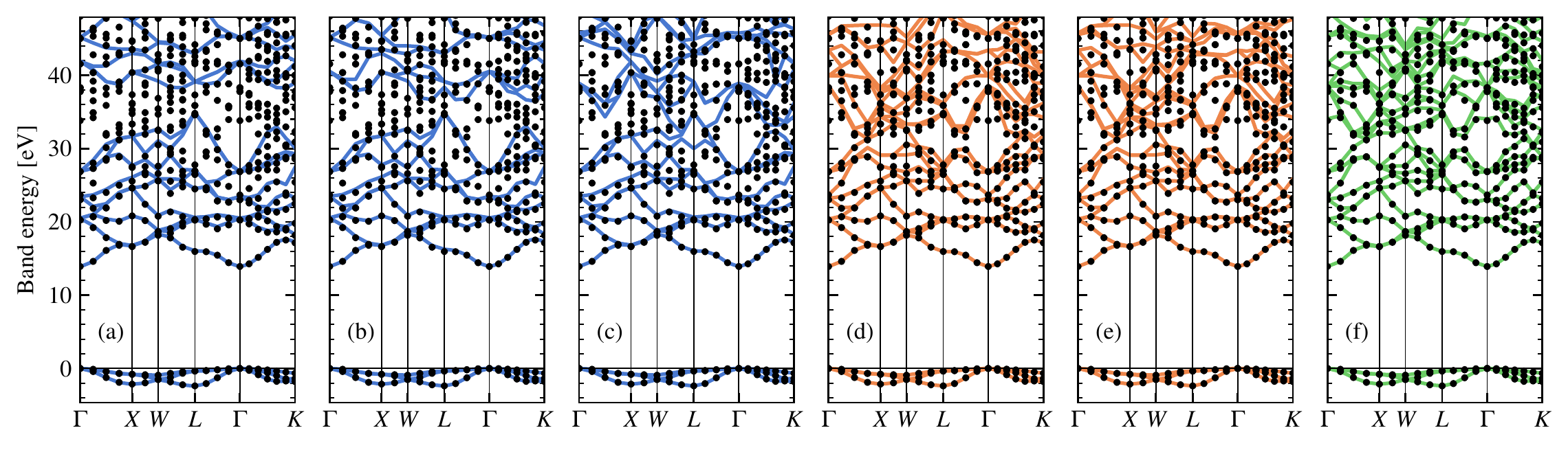}
    \caption{Same plot as \ref{fig:bands_LiH_large} for the band structure of NaCl (large-core).}
    \label{fig:bands_NaCl_large}
\end{figure}

\begin{figure}[h]
    \centering
    \includegraphics[width=1.0\linewidth]{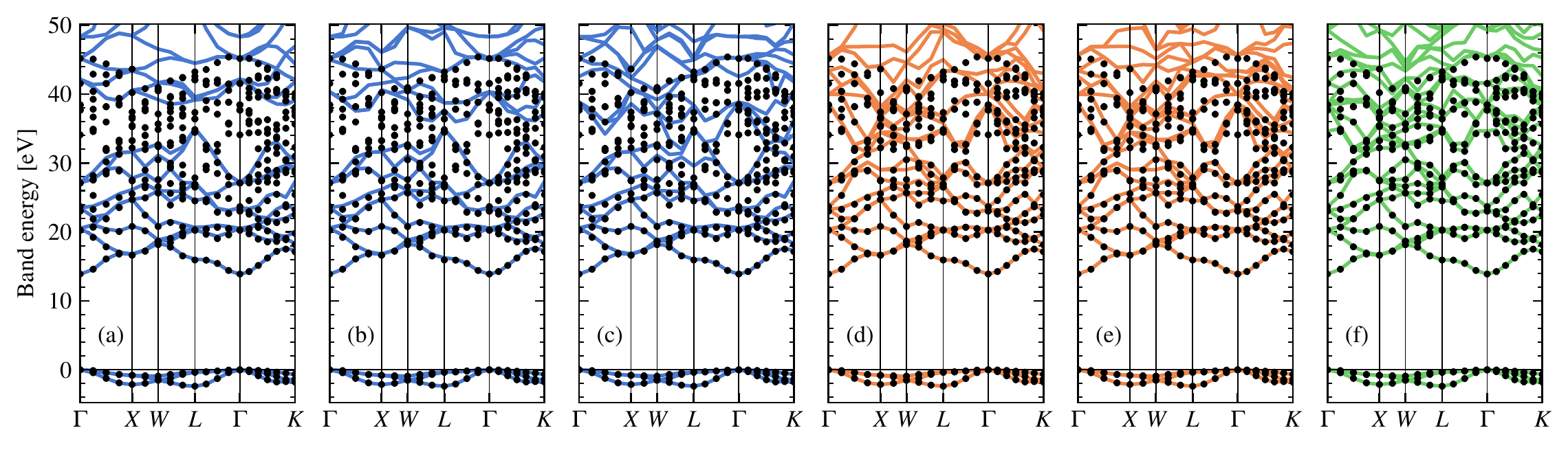}
    \caption{Same plot as \ref{fig:bands_LiH_large} for the band structure of NaCl (small-core).}
    \label{fig:bands_NaCl_small}
\end{figure}

\begin{figure}[h]
    \centering
    \includegraphics[width=1.0\linewidth]{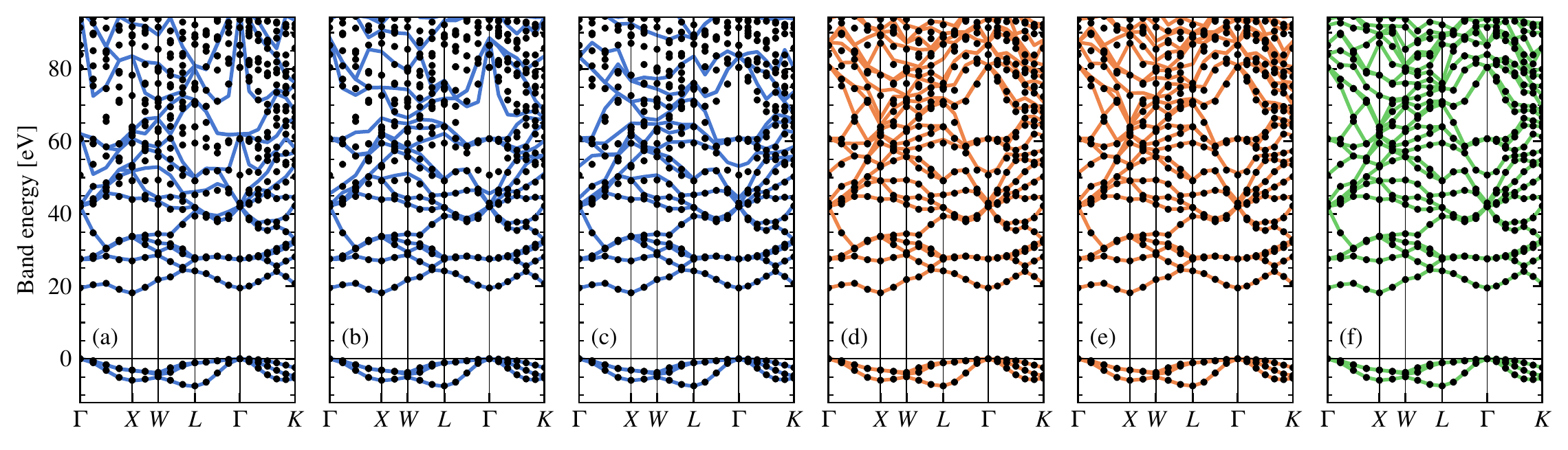}
    \caption{Same plot as \ref{fig:bands_LiH_large} for the band structure of BeO (large-core).}
    \label{fig:bands_BeO_large}
\end{figure}

\begin{figure}[h]
    \centering
    \includegraphics[width=1.0\linewidth]{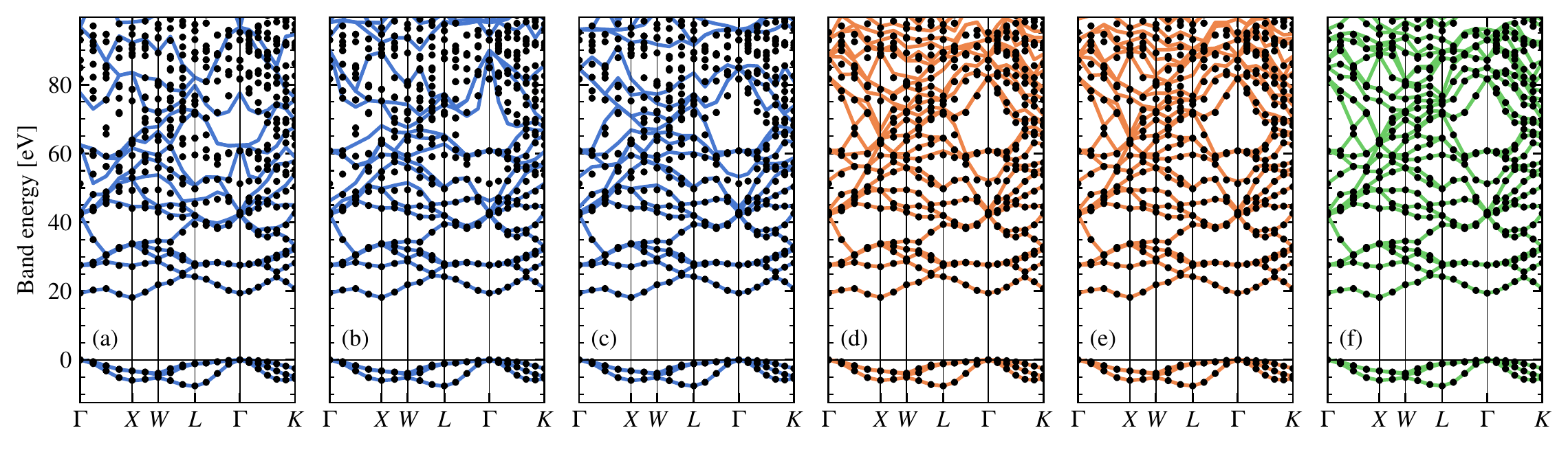}
    \caption{Same plot as \ref{fig:bands_LiH_large} for the band structure of BeO (small-core).}
    \label{fig:bands_BeO_small}
\end{figure}

\begin{figure}[h]
    \centering
    \includegraphics[width=1.0\linewidth]{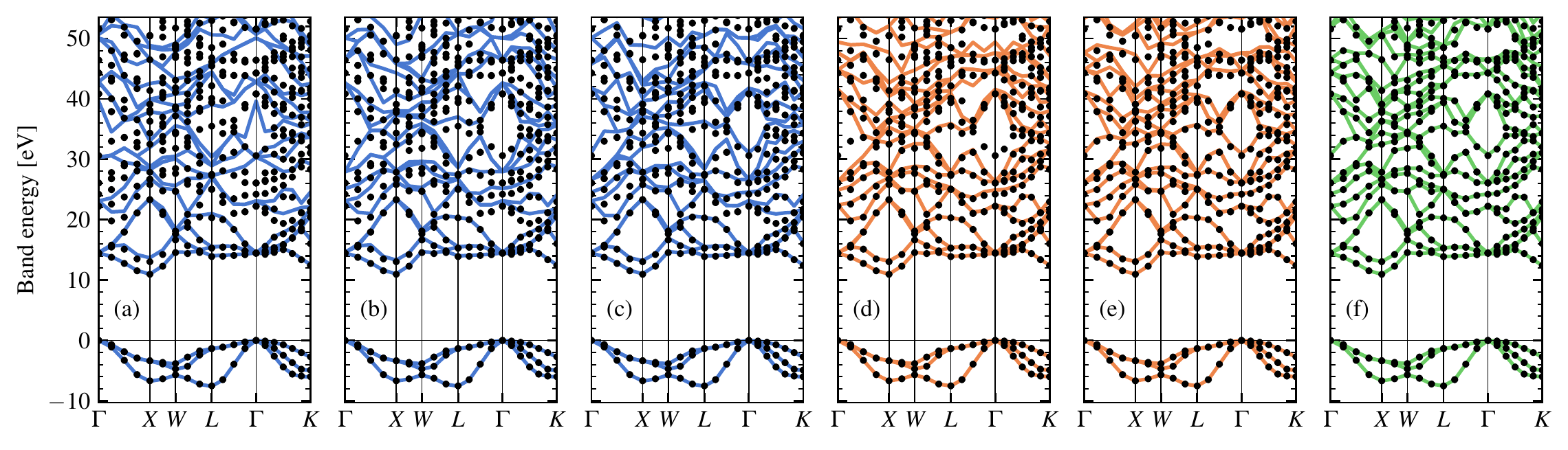}
    \caption{Same plot as \ref{fig:bands_LiH_large} for the band structure of BeS (large-core).}
    \label{fig:bands_BeS_large}
\end{figure}

\begin{figure}[h]
    \centering
    \includegraphics[width=1.0\linewidth]{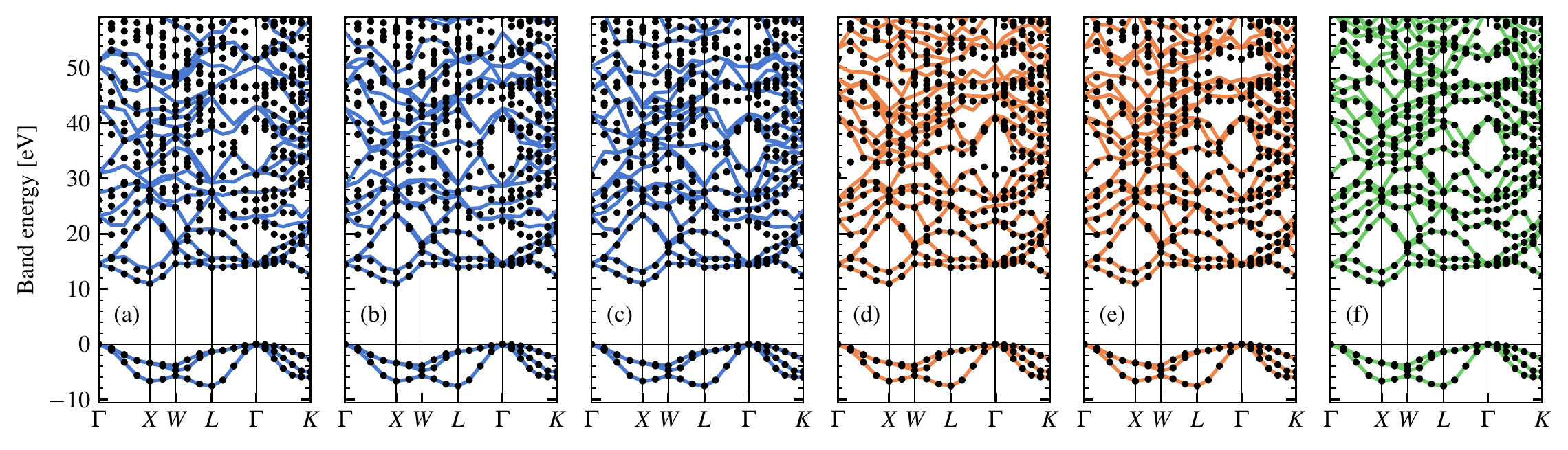}
    \caption{Same plot as \ref{fig:bands_LiH_large} for the band structure of BeS (small-core).}
    \label{fig:bands_BeS_small}
\end{figure}

\begin{figure}[h]
    \centering
    \includegraphics[width=1.0\linewidth]{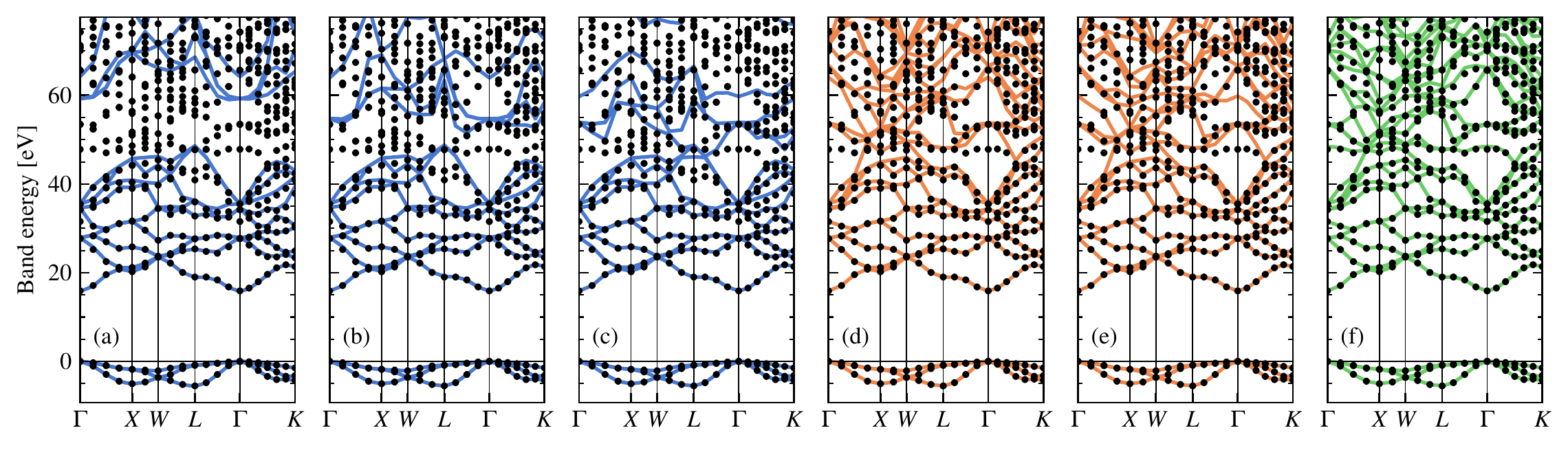}
    \caption{Same plot as \ref{fig:bands_LiH_large} for the band structure of MgO (large-core). The missing state discussed in the main text lies about $30$ eV above the valence band maximum at the $\Gamma$ point.}
    \label{fig:bands_MgO_large}
\end{figure}

\begin{figure}[h]
    \centering
    \includegraphics[width=1.0\linewidth]{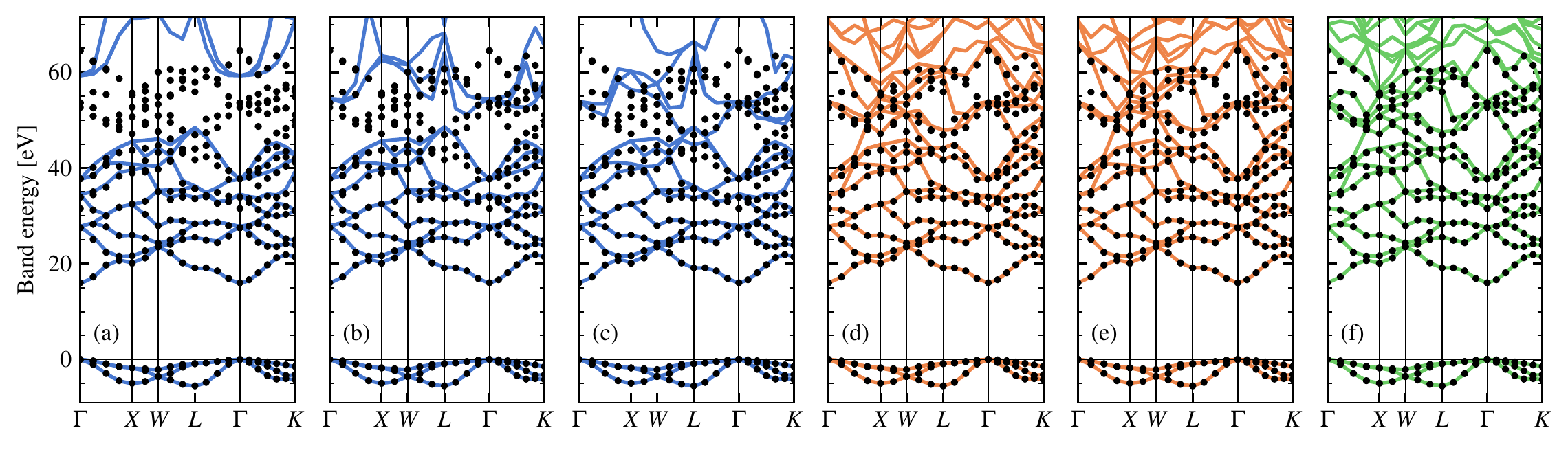}
    \caption{Same plot as \ref{fig:bands_LiH_large} for the band structure of MgO (small-core). The missing state discussed in the main text lies about $30$ eV above the valence band maximum at the $\Gamma$ point.}
    \label{fig:bands_MgO_small}
\end{figure}

\begin{figure}[h]
    \centering
    \includegraphics[width=1.0\linewidth]{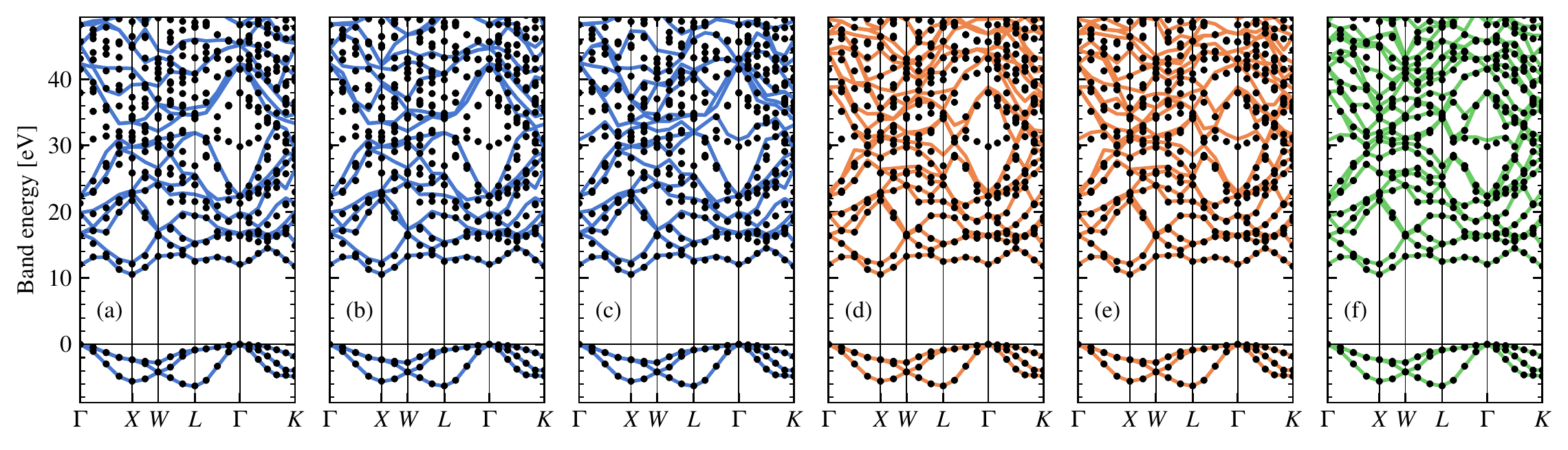}
    \caption{Same plot as \ref{fig:bands_LiH_large} for the band structure of MgS (large-core).}
    \label{fig:bands_MgS_large}
\end{figure}

\begin{figure}[h]
    \centering
    \includegraphics[width=1.0\linewidth]{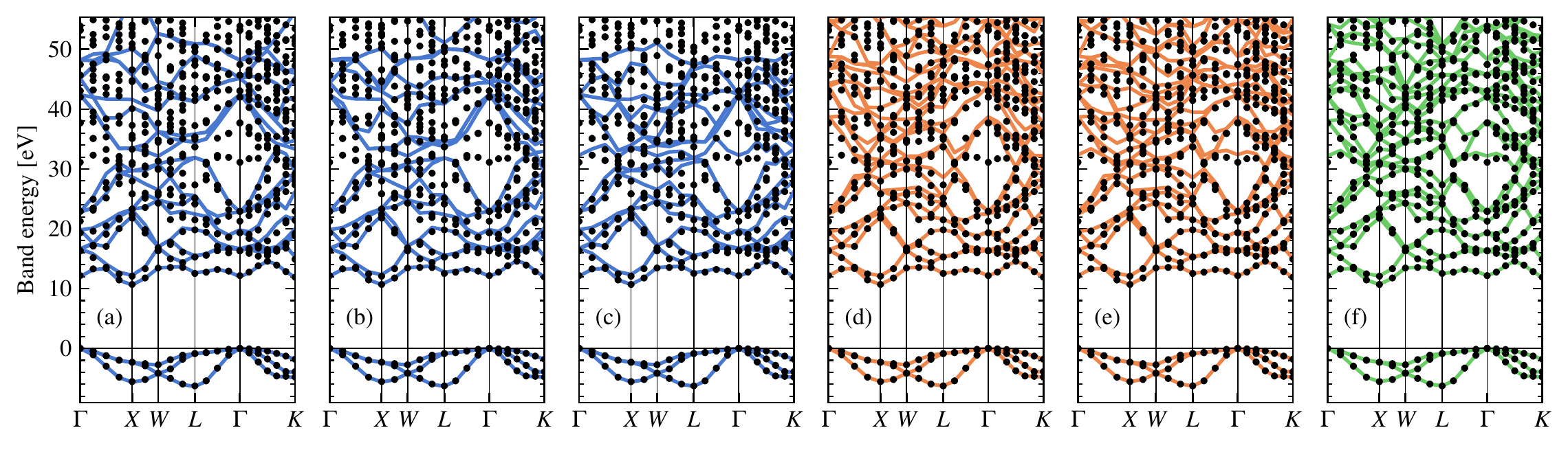}
    \caption{Same plot as \ref{fig:bands_LiH_large} for the band structure of MgS (small-core).}
    \label{fig:bands_MgS_small}
\end{figure}

\begin{figure}[h]
    \centering
    \includegraphics[width=1.0\linewidth]{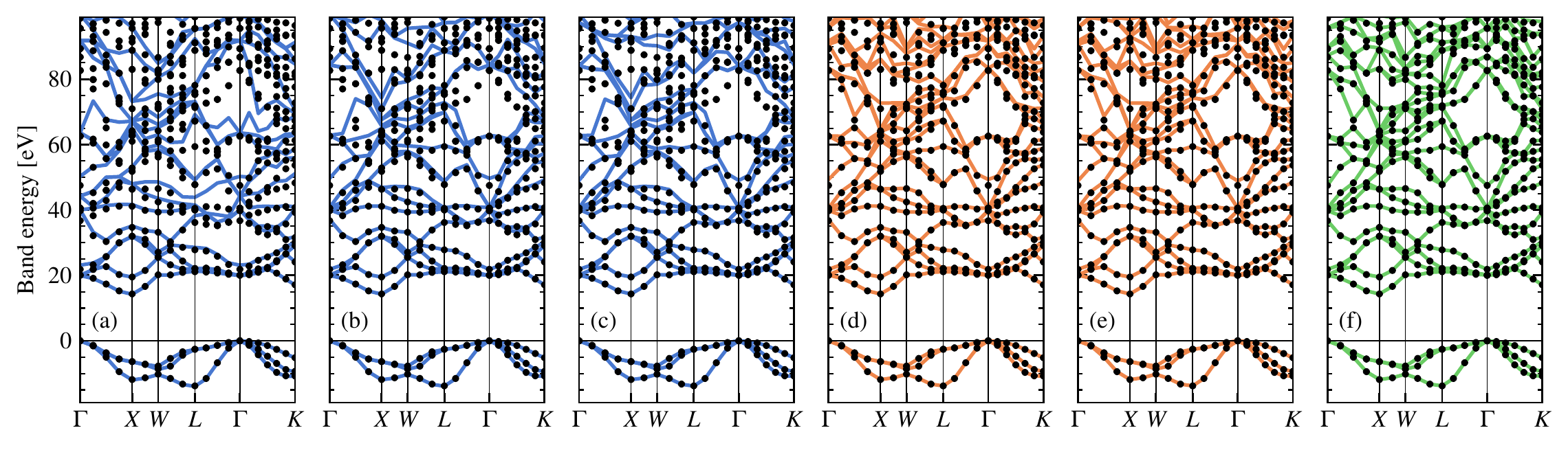}
    \caption{Same plot as \ref{fig:bands_LiH_large} for the band structure of BN.}
    \label{fig:bands_BN}
\end{figure}

\begin{figure}[h]
    \centering
    \includegraphics[width=1.0\linewidth]{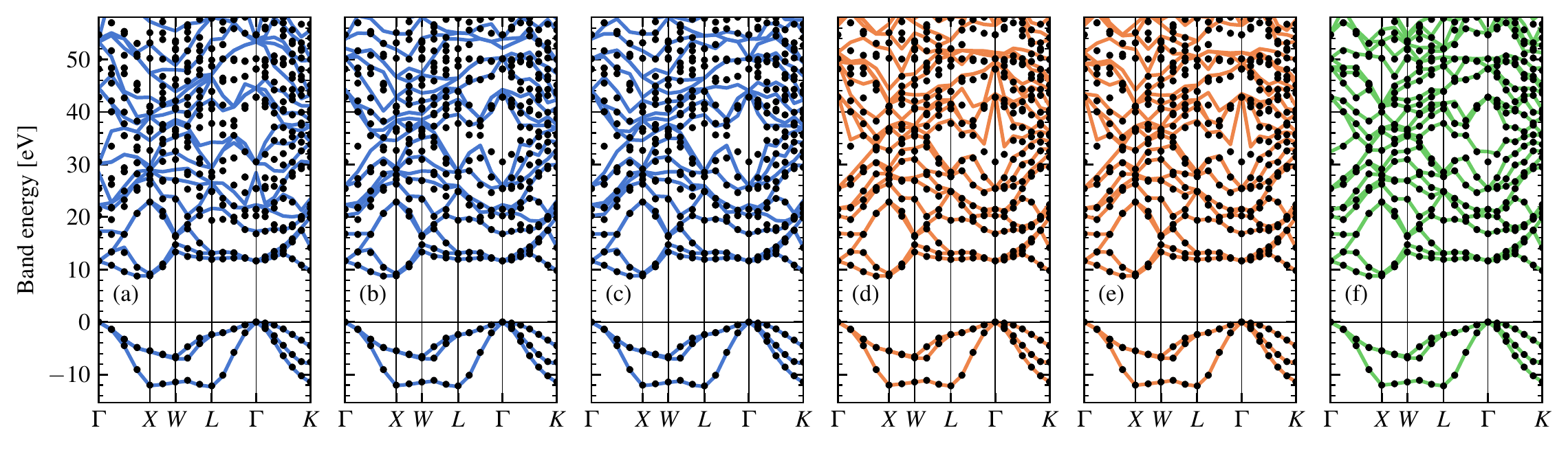}
    \caption{Same plot as \ref{fig:bands_LiH_large} for the band structure of BP.}
    \label{fig:bands_BP}
\end{figure}

\begin{figure}[h]
    \centering
    \includegraphics[width=1.0\linewidth]{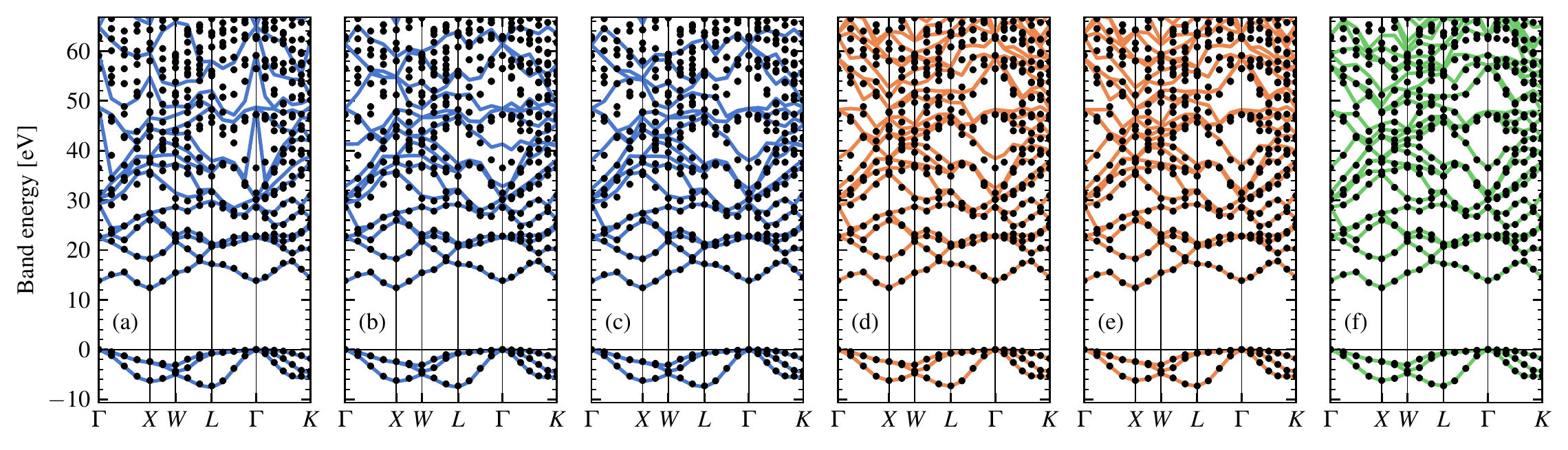}
    \caption{Same plot as \ref{fig:bands_LiH_large} for the band structure of AlN.}
    \label{fig:bands_AlN}
\end{figure}

\begin{figure}[h]
    \centering
    \includegraphics[width=1.0\linewidth]{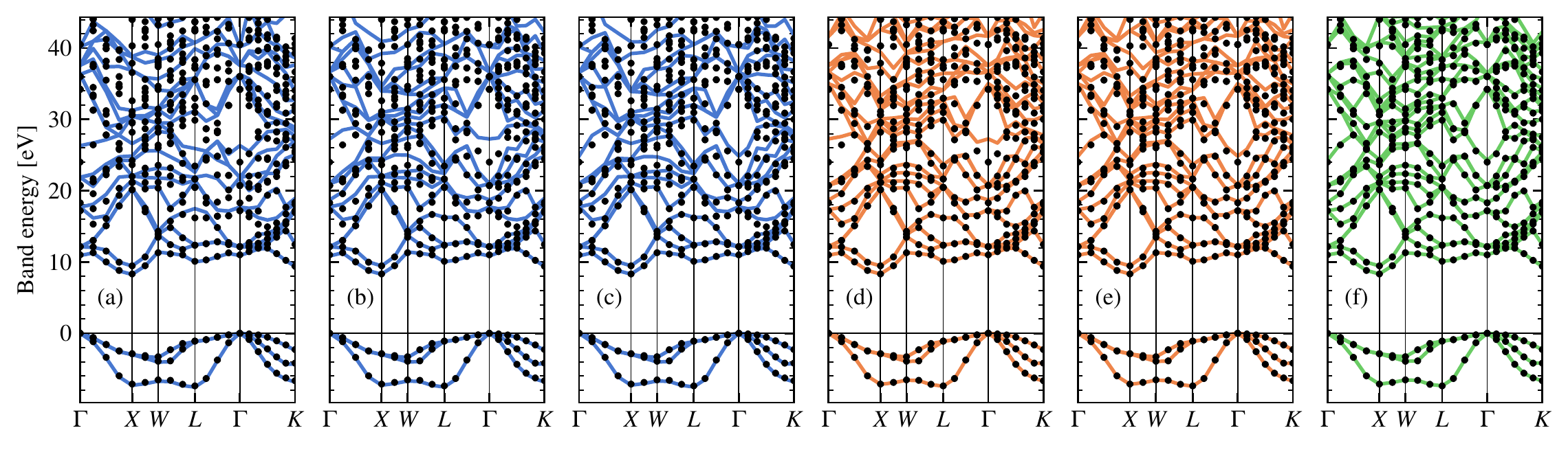}
    \caption{Same plot as \ref{fig:bands_LiH_large} for the band structure of AlP.}
    \label{fig:bands_AlP}
\end{figure}

\begin{figure}[h]
    \centering
    \includegraphics[width=1.0\linewidth]{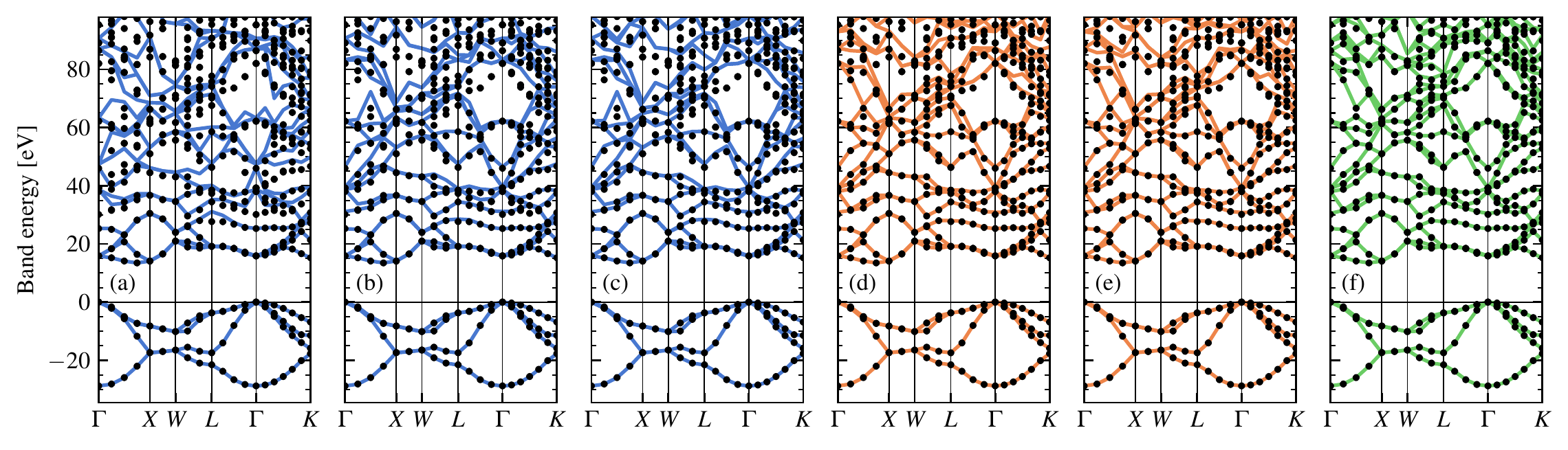}
    \caption{Same plot as \ref{fig:bands_LiH_large} for the band structure of C. Panels (a), (b), (d), and (f) correspond to Fig.~M4(b-e), respectively.}
    \label{fig:bands_C}
\end{figure}

\begin{figure}[h]
    \centering
    \includegraphics[width=1.0\linewidth]{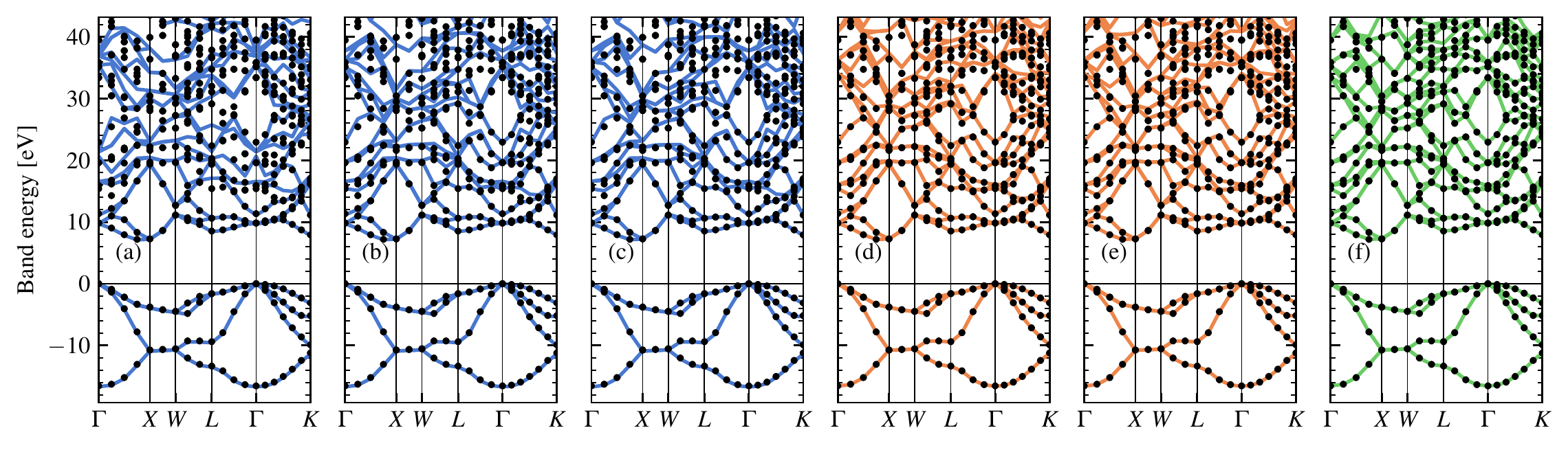}
    \caption{Same plot as \ref{fig:bands_LiH_large} for the band structure of Si.}
    \label{fig:bands_Si}
\end{figure}

\begin{figure}[h]
    \centering
    \includegraphics[width=1.0\linewidth]{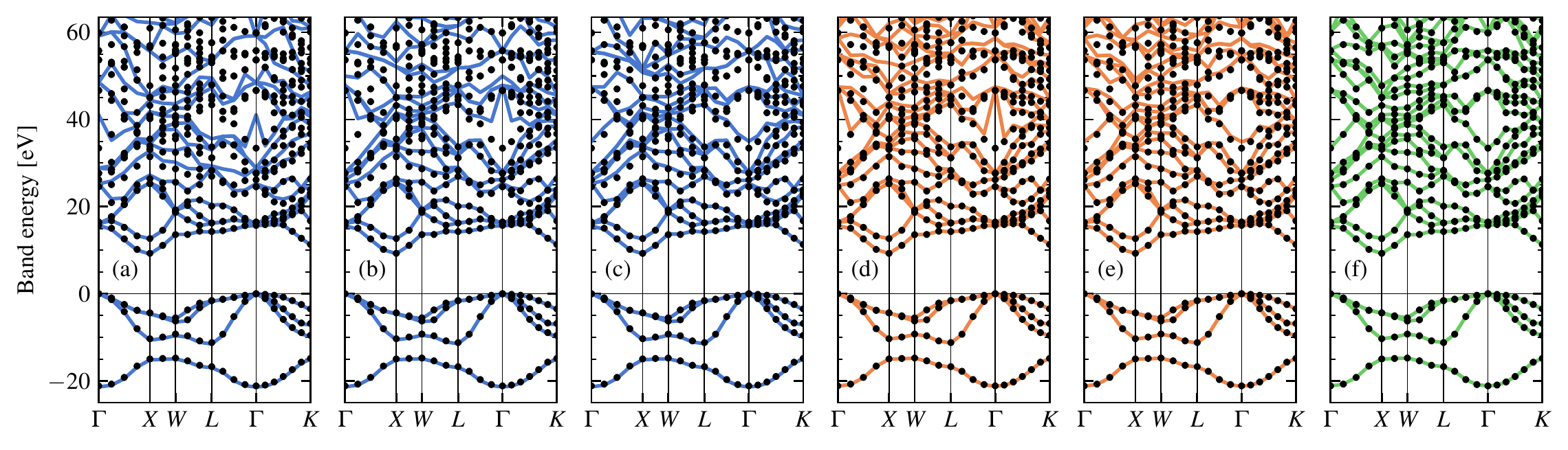}
    \caption{Same plot as \ref{fig:bands_LiH_large} for the band structure of SiC.}
    \label{fig:bands_SiC}
\end{figure}

\begin{figure}[h]
    \centering
    \includegraphics[width=1.0\linewidth]{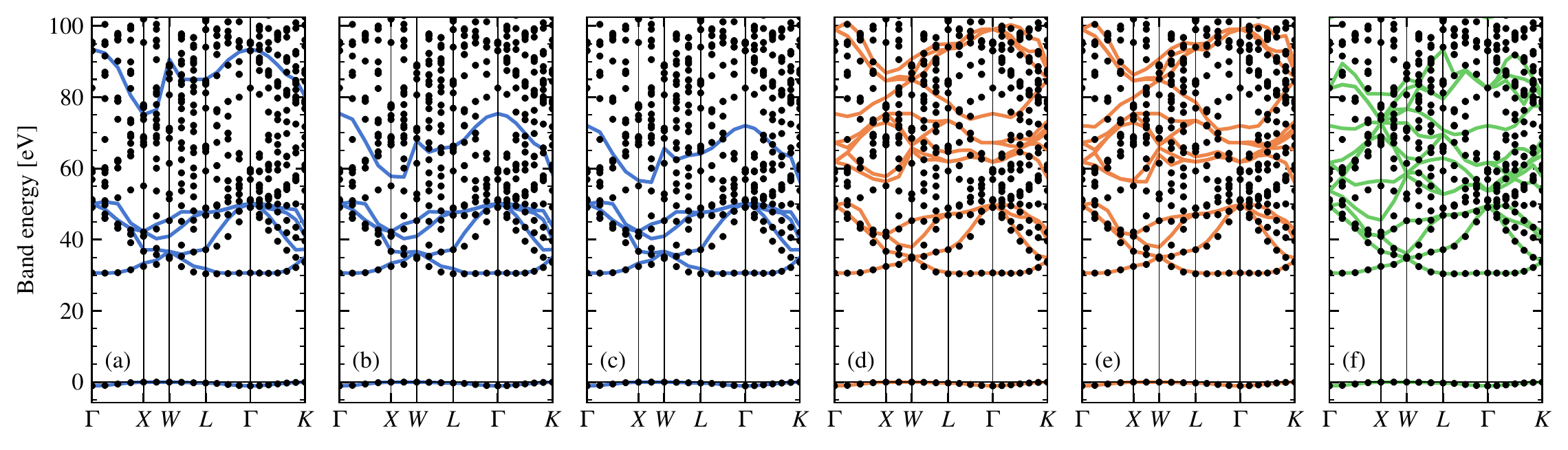}
    \caption{Same plot as \ref{fig:bands_LiH_large} for the band structure of He.}
    \label{fig:bands_He}
\end{figure}

\begin{figure}[h]
    \centering
    \includegraphics[width=1.0\linewidth]{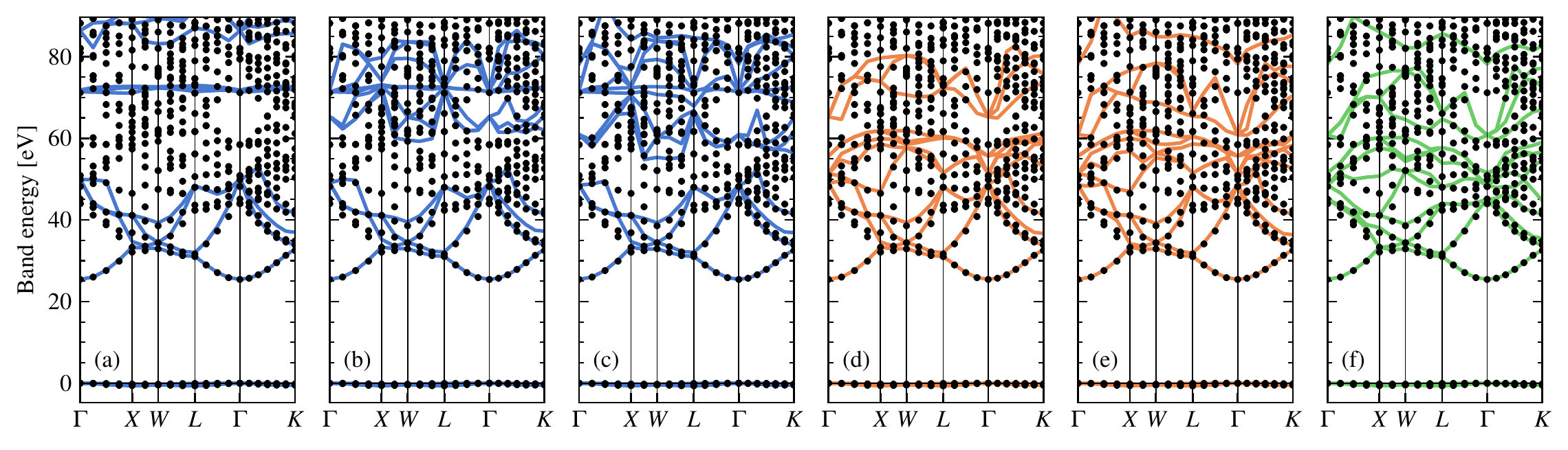}
    \caption{Same plot as \ref{fig:bands_LiH_large} for the band structure of Ne.}
    \label{fig:bands_Ne}
\end{figure}

\begin{figure}[h]
    \centering
    \includegraphics[width=1.0\linewidth]{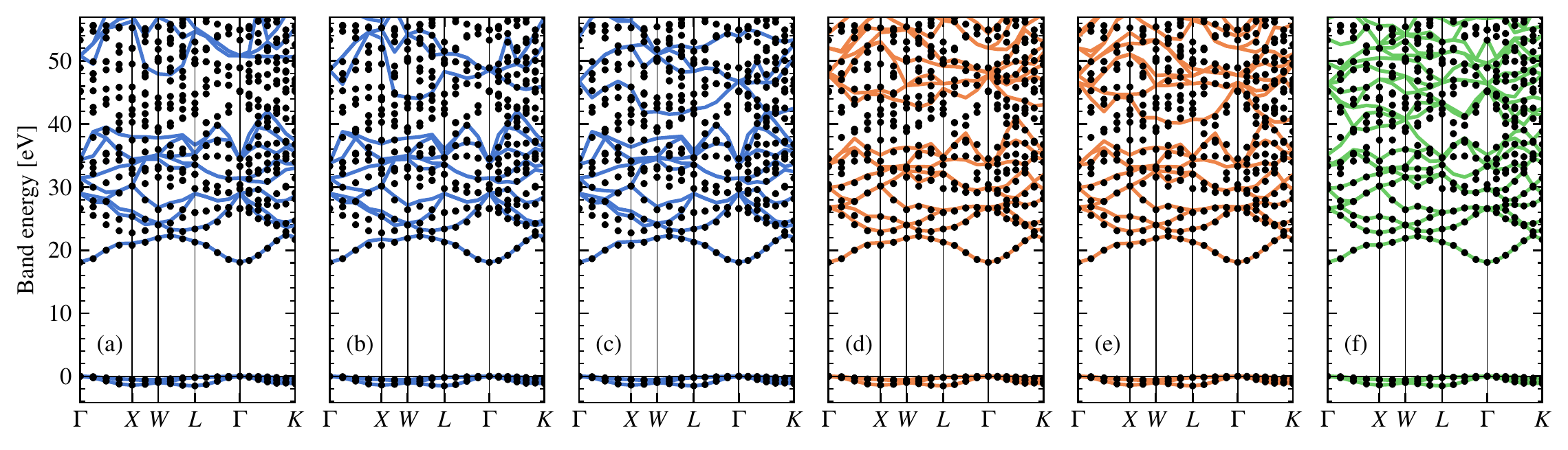}
    \caption{Same plot as \ref{fig:bands_LiH_large} for the band structure of Ar.}
    \label{fig:bands_Ar}
\end{figure}

    \clearpage

    \section{Supplementary tables}

    \begin{longtable}{lllllll}
    \caption{Information of the GTH-cc-pV$X$Z ($X = $ D, T, and Q) basis sets optimized in this work.
    The number of primitive and contracted GTOs are calculated by assuming spherical harmonic functions (i.e.,~$2l+1$ angular components for angualr momentum $l$).} \\
    \hline\hline
    Element & active electrons & zeta-level & pGTOs & $N_{\mathrm{pGTO}}$ & cGTOs & $N_{\mathrm{cGTO}}$ \\
    \hline
    \multirow{3}*{H} & \multirow{3}*{$1s^1$} & GTH-cc-pVDZ & (4s,1p) & $7$ & [2s,1p] & $5$ \\
    & & GTH-cc-pVTZ & (4s,2p,1d) & $15$ & [3s,2p,1d] & $14$ \\
    & & GTH-cc-pVQZ & (4s,3p,2d,1f) & $30$ & [4s,3p,2d,1f] & $30$ \\
    \hline
    \multirow{3}*{He} & \multirow{3}*{$1s^2$} & GTH-cc-pVDZ & (6s,1p) & $9$ & [2s,1p] & $5$ \\
    & & GTH-cc-pVTZ & (7s,2p,1d) & $18$ & [3s,2p,1d] & $14$ \\
    & & GTH-cc-pVQZ & (8s,3p,2d,1f) & $34$ & [4s,3p,2d,1f] & $30$ \\
    \hline
    \multirow{3}*{Li} & \multirow{3}*{$2s^1$} & GTH-cc-pVDZ & (2s,2p,1d) & $13$ & [2s,2p,1d] & $13$ \\
    & & GTH-cc-pVTZ & (2s,2p,1d,1f) & $20$ & [2s,2p,1d,1f] & $20$ \\
    & & GTH-cc-pVQZ & (2s,2p,1d,1f,1g) & $29$ & [2s,2p,1d,1f,1g] & $29$ \\
    \hline
    \multirow{3}*{Li} & \multirow{3}*{$1s^2 2s^1$} & GTH-cc-pVDZ & (4s,4p,1d) & $21$ & [3s,2p,1d] & $14$ \\
    & & GTH-cc-pVTZ & (4s,4p,1d,1f) & $28$ & [4s,3p,1d,1f] & $25$ \\
    & & GTH-cc-pVQZ & (4s,4p,1d,1f,1g) & $37$ & [4s,4p,1d,1f,1g] & $37$ \\
    \hline
    \multirow{3}*{Be} & \multirow{3}*{$2s^2$} & GTH-cc-pVDZ & (3s,3p,1d) & $17$ & [2s,2p,1d] & $13$ \\
    & & GTH-cc-pVTZ & (3s,3p,2d,1f) & $29$ & [3s,3p,2d,1f] & $29$ \\
    & & GTH-cc-pVQZ & (3s,3p,3d,2f,1g) & $50$ & [3s,3p,3d,2f,1g] & $50$ \\
    \hline
    \multirow{3}*{Be} & \multirow{3}*{$1s^2 2s^2$} & GTH-cc-pVDZ & (5s,4p,1d) & $22$ & [3s,2p,1d] & $14$ \\
    & & GTH-cc-pVTZ & (5s,4p,2d,1f) & $34$ & [4s,3p,2d,1f] & $30$ \\
    & & GTH-cc-pVQZ & (5s,4p,3d,2f,1g) & $55$ & [5s,4p,3d,2f,1g] & $55$ \\
    \hline
    \multirow{3}*{B} & \multirow{3}*{$2s^2 2p^1$} & GTH-cc-pVDZ & (3s,3p,1d) & $17$ & [2s,2p,1d] & $13$ \\
    & & GTH-cc-pVTZ & (3s,3p,2d,1f) & $29$ & [3s,3p,2d,1f] & $29$ \\
    & & GTH-cc-pVQZ & (3s,3p,3d,2f,1g) & $50$ & [3s,3p,3d,2f,1g] & $50$ \\
    \hline
    \multirow{3}*{C} & \multirow{3}*{$2s^2 2p^2$} & GTH-cc-pVDZ & (4s,4p,1d) & $21$ & [2s,2p,1d] & $13$ \\
    & & GTH-cc-pVTZ & (4s,4p,2d,1f) & $33$ & [3s,3p,2d,1f] & $29$ \\
    & & GTH-cc-pVQZ & (4s,4p,3d,2f,1g) & $54$ & [4s,4p,3d,2f,1g] & $54$ \\
    \hline
    \multirow{3}*{N} & \multirow{3}*{$2s^2 2p^3$} & GTH-cc-pVDZ & (5s,5p,1d) & $25$ & [2s,2p,1d] & $13$ \\
    & & GTH-cc-pVTZ & (5s,5p,2d,1f) & $37$ & [3s,3p,2d,1f] & $29$ \\
    & & GTH-cc-pVQZ & (5s,5p,3d,2f,1g) & $58$ & [4s,4p,3d,2f,1g] & $54$ \\
    \hline
    \multirow{3}*{O} & \multirow{3}*{$2s^2 2p^4$} & GTH-cc-pVDZ & (6s,6p,1d) & $29$ & [2s,2p,1d] & $13$ \\
    & & GTH-cc-pVTZ & (7s,7p,2d,1f) & $45$ & [3s,3p,2d,1f] & $29$ \\
    & & GTH-cc-pVQZ & (5s,5p,3d,2f,1g) & $58$ & [4s,4p,3d,2f,1g] & $54$ \\
    \hline
    \multirow{3}*{F} & \multirow{3}*{$2s^2 2p^5$} & GTH-cc-pVDZ & (6s,6p,1d) & $29$ & [2s,2p,1d] & $13$ \\
    & & GTH-cc-pVTZ & (7s,7p,2d,1f) & $45$ & [3s,3p,2d,1f] & $29$ \\
    & & GTH-cc-pVQZ & (5s,5p,3d,2f,1g) & $58$ & [4s,4p,3d,2f,1g] & $54$ \\
    \hline
    \multirow{3}*{Ne} & \multirow{3}*{$2s^2 2p^6$} & GTH-cc-pVDZ & (7s,7p,1d) & $33$ & [2s,2p,1d] & $13$ \\
    & & GTH-cc-pVTZ & (8s,8p,2d,1f) & $49$ & [3s,3p,2d,1f] & $29$ \\
    & & GTH-cc-pVQZ & (9s,9p,3d,2f,1g) & $74$ & [4s,4p,3d,2f,1g] & $54$ \\
    \hline
    \multirow{3}*{Na} & \multirow{3}*{$3s^1$} & GTH-cc-pVDZ & (2s,2p,1d) & $13$ & [2s,2p,1d] & $13$ \\
    & & GTH-cc-pVTZ & (2s,2p,1d,1f) & $20$ & [2s,2p,1d,1f] & $20$ \\
    & & GTH-cc-pVQZ & (2s,2p,1d,1f,1g) & $29$ & [2s,2p,1d,1f,1g] & $29$ \\
    \hline
    \multirow{3}*{Na} & \multirow{3}*{$2s^2 2p^6 3s^1$} & GTH-cc-pVDZ & (5s,6p,1d) & $28$ & [3s,2p,1d] & $14$ \\
    & & GTH-cc-pVTZ & (5s,6p,1d,1f) & $35$ & [4s,3p,1d,1f] & $25$ \\
    & & GTH-cc-pVQZ & (5s,6p,1d,1f,1g) & $44$ & [5s,4p,1d,1f,1g] & $38$ \\
    \hline
    \multirow{3}*{Mg} & \multirow{3}*{$3s^2$} & GTH-cc-pVDZ & (2s,1p,1d) & $10$ & [2s,1p,1d] & $10$ \\
    & & GTH-cc-pVTZ & (2s,1p,2d,1f) & $22$ & [2s,1p,2d,1f] & $22$ \\
    & & GTH-cc-pVQZ & (2s,1p,3d,2f,1g) & $43$ & [2s,1p,3d,2f,1g] & $43$ \\
    \hline
    \multirow{3}*{Mg} & \multirow{3}*{$2s^2 2p^6 3s^2$} & GTH-cc-pVDZ & (4s,5p,1d) & $24$ & [3s,2p,1d] & $14$ \\
    & & GTH-cc-pVTZ & (4s,5p,2d,1f) & $36$ & [4s,3p,2d,1f] & $30$ \\
    & & GTH-cc-pVQZ & (4s,5p,3d,2f,1g) & $57$ & [4s,4p,3d,2f,1g] & $54$ \\
    \hline
    \multirow{3}*{Al} & \multirow{3}*{$3s^2 3p^1$} & GTH-cc-pVDZ & (3s,2p,1d) & $14$ & [2s,2p,1d] & $13$ \\
    & & GTH-cc-pVTZ & (3s,2p,2d,1f) & $26$ & [3s,2p,2d,1f] & $26$ \\
    & & GTH-cc-pVQZ & (3s,2p,3d,2f,1g) & $47$ & [3s,2p,3d,2f,1g] & $47$ \\
    \hline
    \multirow{3}*{Si} & \multirow{3}*{$3s^2 3p^2$} & GTH-cc-pVDZ & (4s,4p,1d) & $21$ & [2s,2p,1d] & $13$ \\
    & & GTH-cc-pVTZ & (4s,4p,2d,1f) & $33$ & [3s,3p,2d,1f] & $29$ \\
    & & GTH-cc-pVQZ & (4s,4p,3d,2f,1g) & $54$ & [4s,4p,3d,2f,1g] & $54$ \\
    \hline
    \multirow{3}*{P} & \multirow{3}*{$3s^2 3p^3$} & GTH-cc-pVDZ & (4s,4p,1d) & $21$ & [2s,2p,1d] & $13$ \\
    & & GTH-cc-pVTZ & (4s,4p,2d,1f) & $33$ & [3s,3p,2d,1f] & $29$ \\
    & & GTH-cc-pVQZ & (4s,4p,3d,2f,1g) & $54$ & [4s,4p,3d,2f,1g] & $54$ \\
    \hline
    \multirow{3}*{S} & \multirow{3}*{$3s^2 3p^4$} & GTH-cc-pVDZ & (6s,5p,1d) & $26$ & [2s,2p,1d] & $13$ \\
    & & GTH-cc-pVTZ & (7s,6p,2d,1f) & $42$ & [3s,3p,2d,1f] & $29$ \\
    & & GTH-cc-pVQZ & (5s,4p,3d,2f,1g) & $55$ & [4s,4p,3d,2f,1g] & $54$ \\
    \hline
    \multirow{3}*{Cl} & \multirow{3}*{$3s^2 3p^5$} & GTH-cc-pVDZ & (6s,5p,1d) & $26$ & [2s,2p,1d] & $13$ \\
    & & GTH-cc-pVTZ & (7s,6p,2d,1f) & $42$ & [3s,3p,2d,1f] & $29$ \\
    & & GTH-cc-pVQZ & (5s,4p,3d,2f,1g) & $55$ & [4s,4p,3d,2f,1g] & $54$ \\
    \hline
    \multirow{3}*{Ar} & \multirow{3}*{$3s^2 3p^6$} & GTH-cc-pVDZ & (5s,5p,1d) & $25$ & [2s,2p,1d] & $13$ \\
    & & GTH-cc-pVTZ & (6s,6p,2d,1f) & $41$ & [3s,3p,2d,1f] & $29$ \\
    & & GTH-cc-pVQZ & (7s,7p,3d,2f,1g) & $66$ & [4s,4p,3d,2f,1g] & $54$ \\
    \hline
\end{longtable}

    \begin{table}
        \centering
        \caption{The 19 materials studied in this work. The listed lattice constants are for the eight-atom cubic cell while the two-atom primitive cells are used in all our simulations, except for the noble gas solids, where the listed lattice constants are for the four-atom cubic cell while the one-atom primitive cells are used in our simulations.}
        \begin{tabular}{lll}
            \hline\hline
            formula & lattice type & $a_0 / \mathrm{\AA}$ \\
            \hline
            LiH & rocksalt & $4.083$ \\
            LiF & rocksalt & $4.035$ \\
            LiCl & rocksalt & $5.130$ \\
            NaF & rocksalt & $4.620$ \\
            NaCl & rocksalt & $5.640$ \\
            BeO & zincblende & $3.797$ \\
            BeS & zincblende & $4.870$ \\
            MgO & rocksalt & $4.207$ \\
            MgS & rocksalt & $5.200$ \\
            BN & zincblende & $3.616$ \\
            BP & zincblende & $4.538$ \\
            AlN & zincblende & $4.380$ \\
            AlP & zincblende & $5.463$ \\
            C & diamond & $3.567$ \\
            Si & diamond & $5.430$ \\
            SiC & zincblende & $4.358$ \\
            He & fcc & $4.112$ \\
            Ne & fcc & $4.446$ \\
            Ar & fcc & $5.311$ \\
            \hline
        \end{tabular}
    \end{table}